\documentclass[useAMS,usenatbib]{mn2e}
\usepackage{amssymb,amsmath,epsfig,times, natbib}
\def\keV{{\rm\thinspace keV}}

\title[The outskirts of Centaurus]{X-ray exploration of the outskirts of the nearby Centaurus cluster using \emph{Suzaku} and \emph{Chandra}}
\author[S. A. Walker et al.]{S. A. Walker,$^1$\thanks{Email: 
    swalker@ast.cam.ac.uk} A. C. Fabian,$^1$ J. S. Sanders$^2$, A. Simionescu$^{3,4}$ and Y. Tawara$^5$\\
  $^1$Institute of Astronomy, Madingley Road, Cambridge CB3 0HA \\
  $^2$Max-Planck-Institute fur extraterrestrische Physik, 85748 Garching, Germany \\
  $^3$KIPAC, Stanford University, 452 Lomita Mall, Stanford, CA 94305, USA\\
  $^4$Department of Physics, Stanford University, 382 Via Pueblo Mall, Stanford, CA 94305-4060, USA\\
    $^5$Department of Physics, Nagoya University, Nagoya, 338-8570, Japan 
}
\date{}

\begin{document}

\maketitle

\begin{abstract}
We present Suzaku observations of the Centaurus cluster out to 0.95$r_{200}$, taken along a strip to the north west. We have also used congruent Chandra observations of the outskirts to resolve point sources down to a threshold flux around 7 times lower than that achievable with just Suzaku data, considerably reducing the systematic uncertainties in the cosmic X-ray background emission in the outskirts. We find that the temperature decreases by a factor of 2 from the peak temperature to the outskirts. The entropy profile demonstrates a central excess (within 0.5$r_{200}$) over the baseline entropy profile predicted by simulations of purely gravitational hierarchical structure formation. In the outskirts the entropy profile is in reasonable agreement with the baseline entropy profile from Voit et al., but lies slightly below it. We find that the pressure profile agrees with the universal pressure profile of Arnaud et al. but lies slightly above it in the outskirts. The excess pressure and decrement in entropy in the outskirts appear to be the result of an excess in the measured gas density, possible due to gas clumping biasing the density measurements high. The gas mass fraction rises and reaches the mean cosmic baryon fraction at the largest radius studied. The clumping corrected gas mass fraction agrees with the expected hot gas fraction and with the simulations of Young et al. We further the analysis of Walker et al. which studied the shapes of the entropy profiles of the clusters so far explored in the outskirts with Suzaku. When scaled by the self similar entropy the Suzaku entropy profiles demonstrate a central excess over the baseline entropy profile, and are consistent with it at around $r_{500}$. However outside $r_{500}$ the entropy profiles tend to lie below the baseline entropy profile.
 
\end{abstract}


\section{Introduction}
The low and stable particle background of Suzaku has allowed significant progress to be made in the study of the low surface brightness X-ray emission from galaxy cluster outskirts, providing exciting and unique observations of these previously unexplored regions. In the outskirts, gas is continuing to accrete onto clusters, allowing us to see the formation process in action. We can gain an understanding of how close the ICM is to hydrostatic equilibrium and spherical symmetry in the outskirts, which are the fundamental assumptions made when deriving masses of galaxy clusters from X-ray observations. Estimations of the cluster mass within $r_{200}$\footnote{r$_{200}$ is the radius within which the mean density of the cluster is 200 times the critical density required for a flat universe, $\rho_{c}$, and is typically used to represent the virial radius. The mass enclosed within r$_{200}$ is M$_{200}$=4/3$\pi\rho_{c}$r$_{200}^{3}$. } have previously involved extrapolating outwards analytic best fit temperature and density profiles for data from within $r_{500}$ $\sim$ $\frac{2}{3}r_{200}$, (for example \citealt{Vikhlinin2006}), meaning that only $\sim$ 30 percent of the cluster volume had actually been explored. This is subject to large errors and priors, as there is no robust prior expectation for the form of the temperature and density profiles in the outskirts. 

Accurate cluster mass determinations are important for using clusters as probes of cosmological parameters using the mass function method (\citealt{Vikhlinin2006}, \citealt{Mantz2010}). Improving our understanding of the gas mass fraction of clusters at large radius allows us to better understand the assumptions behind using the gas mass fraction as a cosmological probe \citep{Allen2008}. 

The breakthrough in the study of cluster outskirts with Suzaku was made for clusters whose size and redshift allowed the virial radius to be reached using a small number of pointings, in that the angular extent of the cluster made the distance between the core and $r_{200}$ around 20-30 arcmins. These clusters are PKS 0745-191 \citep{George2009, Walker2012_PKS0745}, Abell 2204 \citep{Reiprich2009}, Abell 1795 \citep{Bautz2009}, Abell 1413 \citep{Hoshino2010}, Abell 1689 \citep{Kawaharada2010}, Abell 2142 \citep{Akamatsu2011}, the fossil group RX J1159+5531 \citep{Humphrey2011}, Hydra A \citep{Sato2012} and Abell 2029 \citep{Walker2012_A2029}. However the large point spread function (PSF) of Suzaku (HPD=2$'$) limits the size of annuli used in spatially resolved spectral analysis, reducing the resolution of the temperature and entropy profiles.

To get around the large PSF and improve the spatial resolution of the profiles of the ICM properties, we need to observe nearby, X-ray bright clusters, as was achieved in \citet{Simionescu2011} and \citet{Simionescu2012} for the Perseus cluster. The Centaurus cluster's low redshift (z=0.0109) gives an angular extent of 13 kpc/arcmin, meaning more spatially resolved temperature and entropy profiles can be obtained. Centaurus also lies further from the galactic plane than Perseus giving it a lower absorbing column and reducing the fluctuations in column density across the face of cluster. 

At present, Suzaku has mainly studied high mass clusters to the virial radius [the fossil group RX J1159+5531, Hydra A (3keV) and Virgo (2.3keV) are the only clusters below 4 keV to be studied to $r_{200}$]. Centaurus' low average temperature (we find $kT(0.1<r<0.5r_{200})=$ 3.0$_{-0.1}^{+0.1}$ keV) means that we are probing the low mass range of clusters, and so its study is of importance as it will allow us to gain an understanding of the properties of clusters of different masses in the outskirts. It is important to explore the full mass range of clusters to investigate $M-T$, $M-L$ and $Y_{X}-M$ scaling relations throughout the complete mass range. For instance, \citet{Arnaud2005} found that when clusters below  $kT(0.1<r<0.5r_{200})=$ 3.5 keV are included in the M-T relation, it steepens compared to the self-similar relation $M \varpropto T^{1.5}$, becoming $M \varpropto T^{1.7}$. 

Of particular interest is the entropy profile ($K=kT/n_{e}^{2/3}$). Numerical simulations of pure gravitational collapse \citep{Voit2005} predict that the entropy should increase as a powerlaw with radius ($K \varpropto r^{1.1}$) and also provides a prediction for the normalisation of this baseline entropy profile. Deviations from this baseline entropy profile must originate from non-gravitational processes, and so comparing the shape and normalisation of the observed entropy profile with the baseline entropy profile provides an insight into the physical processes occurring in the ICM. 

We have already shown in \citet{Walker2012_UEP} that the entropy profiles for clusters explored with Suzaku out to $r_{200}$ have the same shape, flattening away from a powerlaw increase above $\sim$0.5$r_{500}$. The same entropy profile shape was also found for the XMM-Newton study of the Virgo cluster \citep{Urban2011}, and the Chandra studies of A2204 \citep{Sanders2009} and A1835 \citep{Bonamente2012}. However, more insight can be obtained by scaling the entropy profiles by the self similar entropy at $r_{500}$, which allows both that shape and normalisation to be compared to the baseline profile. This is explored in section \ref{UEP}. 

A similar entropy analysis has been performed in \citet{Pratt2010} for the entropy profiles of the REXCESS clusters (XMM-Newton observations of 31 clusters extending to at least $r_{1000}$, and up to $r_{500}$ for 13 clusters) and in \citet{Sun2009} for a sample of groups. In both cases the entropy profiles are flatter than the $K \varpropto r^{1.1}$ powerlaw prediction at $r_{500}$. However the entropy is found to exceed the baseline level inside $r_{500}$, such that the flattening only acts to bring the entropy level into agreement with the baseline level at $r_{500}$, without ever requiring it to go below the baseline entropy profile. This entropy excess is possibly the result of a combination of extra heating and mixing of the ICM caused by mergers. The central entropy excess is mass dependent, with less massive clusters showing a greater excess. 

Also of interest is the pressure profile. \citet{Arnaud2010} used the REXCESS sample of clusters to show that the pressure profiles of clusters tend to have the same shape and normalisation when scaled by the characteristic pressure, $P_{500}$, and have used predictions of simulations to extend the proposed universal pressure profile out to and beyond $r_{200}$. Observations with Planck \citep{PlanckV2012} using the Sunyaev-Zeldovich effect to measure the pressure profiles of a stacked sample of 62 clusters have agreed with the form of this universal pressure profile in the outskirts of clusters, but has found slightly higher pressures than simulations predict outside $r_{500}$. 
   
Cluster outskirts observations are also important in constraining the level of gas clumping occurring the ICM, which is a prediction of numerical simulations \citep{Roncarelli2006}. Gas clumping causes the gas density to be overestimated if the ICM is assumed to be uniform, resulting in underestimates of the entropy \citep{Nagai2011}, and overestimates of the gas pressure and gas mass fraction, as has been observed for Perseus \citep{Simionescu2011} and PKS 0745-191 \citep{Walker2012_PKS0745}. Simulations also predict that the contribution of non-thermal pressure support in the ICM increases to around 20 percent at $r_{200}$ \citep{Lau2009}, causing hydrostatic equilibrium masses which use only the gas pressure to underestimate the total cluster mass. 

We can also study the gas mass fraction profile, which depends on cluster mass, and which will allow feedback models \citep{Young2011} to be tested. In less massive clusters it is expected that feedback processes are more able to redistribute gas to larger radius, and these observations provide an excellent way of testing these models. 

Here we present Suzaku observations of the Centuarus cluster, taken along a strip from the core to the outskirts in the north west direction, reaching out to $r_{200}$ and beyond. The north west direction was chosen because it avoids the cold front to the west (studied with Chandra in \citealt{Fabian2005}), and also avoids the merging activity reported in \citet{Churazov1999} to the south east. We therefore expect this direction to be relaxed, allowing a hydrostatic mass analysis to be performed and a determination of the gas mass fraction. 

\begin{figure*}
  \begin{center}
    \leavevmode
      \epsfig{figure=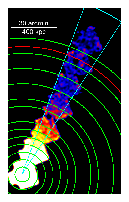,
        width=0.25\linewidth}
       \epsfig{figure=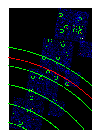,
        width=0.25\linewidth} 
         \epsfig{figure=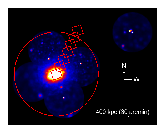,
        width=0.45\linewidth}     
        
      \caption{\emph{Left}:Mosaic of Suzaku pointings in the 0.7-7.0 keV band. The annuli used in spectral extraction are shown in green and the ring radii are at 5,10,15,20,25,30,35,45,50,60,70,80,and 90 arcmins. The value of $r_{200}$ calculated later in the mass analysis is shown by the red line. The cyan sector shows the region which is free from stray light due to the pointings being orientated with the chip diagonal towards the core. Point sources which are resolved in the Suzaku images have been removed from the image (and from the spectral analysis). \emph{Centre}: Mosaic of Chandra ACIS-I observations (0.5-7.0 keV band) of the regions more the 50 arcmins from the cluster centre. Point sources identified with \textsc{wavdetect} are circled. The spectral extraction annuli are the same as shown in the left panel.  \emph{Right}: Mosaic of ROSAT PSPC pointings with the Suzaku pointing locations overlaid in red. The red circle shows the value of $r_{200}$ we calculate later. The offset pointing is of NGC 4507 and is used as a background pointing in the ROSAT analysis we perform. }
      \label{annuliPKS}
  \end{center}
\end{figure*}

\begin{figure}
  \begin{center}
    \leavevmode
    \hbox{
      \epsfig{figure=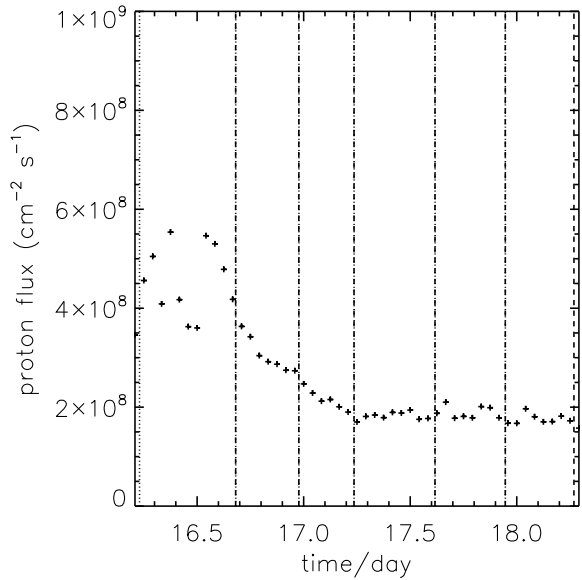,width=\linewidth} 
        }
      \caption{Checking for SWCX contamination during the periods of observation using the  proton flux obtained with the WIND spacecraft. SWCX is important when the proton flux is greater then 4$\times10^{8} cm^{-2}s^{-1}$. Vertical dashed lines show the start and end of each observation. The pointings where taken in order of increasing radius from the centre. The only pointing for which the proton flux exceeds  4$\times10^{8} cm^{-2}s^{-1}$ is the central most pointing (806084010). The fact that the excess is only slight, combined with the high brightness of the cluster in this region near the core, makes the effects of possible SWCX contamination negligible.   }
      \label{ACE}
  \end{center}
\end{figure}

We use a standard $\Lambda$CDM cosmology with $H_{0}=70$  km s$^{-1}$
Mpc$^{-1}$, $\Omega_{M}=0.3$, $\Omega_{\Lambda}$=0.7. All errors unless
otherwise stated are at the 1 $\sigma$ level.

All spectral fits were performed in \textsc{xspec} 12.7.0u using the extended C-statistic.

\section{Observations and Data Reduction}

Six observations were taken in a strip to the north west, avoiding known cold fronts and bright point sources. The roll angle was chosen so that pointing diagonal points towards the core, which has been found to minimise the effects of stray light (as shown in Figs. 6.18 and 6.19 of the Suzaku Technical Description\footnote{heasarc.nasa.gov/docs/suzaku/prop\_tools/suzaku\_td/suzaku\_td.html}) due to the design of the X-ray telescopes (which were built as four sectors). The stray light free region is a sector of angle 12.8 degrees when the chip diagonal points towards the cluster core (see Fig. 6.19 of the Suzaku Technical Description), and this is shown by the cyan sector in Fig. \ref{annuliPKS}, which shows that in the outskirts the pointings are largely free of stray light. We use an archival Suzaku core observation to complete the mosaic image. The observations used are shown in table \ref{obsdetails}, and the mosaic image of these pointings is shown in Fig. \ref{annuliPKS}, left. 

Four 10ks Chandra ACIS-I observations were obtained (PI: S.A. Walker) of the regions covered by the outermost Suzaku pointings (outside 50 arcmins from the core), to better resolve point sources and reduce the systematic errors of the uncertainty of the CXB level on the measurements. The importance of using Chandra to resolve point sources in Suzaku analyses of cluster outskirts is also described in \cite{Miller2012}. The observations are shown in Fig. \ref{annuliPKS} (centre) and the details are shown in table \ref{obsdetails}. 

In addition we use 7 archival ROSAT PSPC observations shown in table \ref{obsdetails} and mosaicked in Fig. \ref{annuliPKS} (right), which are used later in determining the expected spatial variations of the soft galactic background components, and in calculating a density profile to check the Suzaku results.

\begin{table*}
  \begin{center}
  \caption{Observational parameters of the pointings}
  \label{obsdetails}
  
    \leavevmode
    \begin{tabular}{llllllll} \hline \hline
    Instrument & Obs. ID & Position & Total exposure & RA & Dec (J2000) & Date & $N_{H}$ / 10$^{20}$ cm$^{-2}$ \\ 
    & & & per detector (ks)& & & & \citet{LABsurvey}\\ \hline
    Suzaku XIS & 800014010 & Centre & 28.2  & 192.2012 & -41.3132&  2005-12-27& 8.30\\
     & 806084010 & 1 & 10.9  & 192.0509 & -41.0558& 2012-01-16 & 8.43\\
     & 806085010 & 2 & 13.4  & 191.8847 & -40.8272&  2012-01-16 &  8.41\\
     & 806086010 & 3 & 10.0  & 191.7105 & -40.5689&  2012-01-16 &  8.67\\
     & 806087010 & 4 & 10.1  & 191.5735 & -40.3305&  2012-01-17 & 8.05\\
     & 806088010 & 5 & 13.2  & 191.4290 & -40.0920& 2012-01-17 & 8.14\\
     & 806089010 & 6 & 11.1  & 191.2847 & -39.8559&  2012-01-17 & 7.76\\ \hline
    
     Chandra ACIS-I &15182 &  &  10 & 191.675 & -40.448	& 2012-11-26 & \\
     &15183 &  & 10  & 191.488 & 	-40.18& 2012-11-26 &\\
     & 15184&  & 10  & 191.324 & 	-39.91&2012-11-25 & \\
     &15185 &  &  10 &191.1475  & -39.638	&2012-11-25 & \\ \hline
    
     ROSAT PSPC &rp800192n00 &Centre  &  7.793 & 192.2000 & -41.31000 &	& \\
     &rp800607n00 &Centre  & 6.787  & 192.2000 & 	-41.31000 & & \\
     & rp800323n00&North  & 20.658  & 192.5500 & 	-40.52000 & &\\
     &rp800321n00 &East  &  1.584 & 193.2600 & -41.57000	& & \\
     &rp800322n00 &West  &  16.908 & 191.1500 &  -41.06000 &  & \\
     & rp800324n00&South   & 3.782  & 191.8600	 & 	-42.11000 &  &\\ 
     & rp701518n00&NGC 4507   & 5.882  & 188.9000	 & 	-39.91000 & & \\   \hline
    
    \end{tabular}
  \end{center}
\end{table*}

\subsection{Suzaku data reduction}

The Suzaku data were reduced using the method described in \citet{Walker2012_A2029} using HEAsoft version 6.12 and the CALDB released on 2012 October 5, and we used the latest contamination layer calibration. The observations were reprocessed using the ftool \textsc{aepipeline}, which performs the standard cleaning described in tables 6.1 and 6.2 of the Suzaku ABC guide\footnote{http://heasarc.gsfc.nasa.gov/docs/suzaku/analysis/abc/}, and in addition we used the COR$>$6 condition. Spectra were then extracted from the XIS0, XIS1 and XIS3 detectors using the annulus regions shown in Fig. \ref{annuliPKS}. The calibration regions at the corners of the detectors were removed, and the regions at the edges of the detectors where the effective area is low were removed. For the XIS0 detector, the defective region (from a micrometeorite hit) was removed. Outside $r=$ 50 arcmins, the small regions of the Suzaku data which did not overlap with the Chandra data were excluded from the spectral extraction.

For each spectrum the non X-ray background (NXB) spectrum was produced using the ftool \textsc{xisnxbgen} which uses a database of night earth observations, and we used the standard 300 day interval (between 150 days before and 150 days after the observation) to construct the NXB spectra. The latest calibration files were used, which account for the modification of the NXB level of the XIS1 detector due to the increase in the charge injection level on 2011 June 1. We produced two ARFs for each spectrum using \textsc{xissimarfgen} due to the different spatial variations of the background and cluster emission. The first assumes a uniform source and is used for the background model when performing spectral fitting in xspec. The second used a background subtracted image of the cluster, and is used for the model of the cluster emission. 

The light curves were checked to ensure that no flaring occurred during the remaining cleaned observations. We then checked that the observations were not contaminated by solar wind charge exchange emission (SWCX) by investigating the proton flux from the WIND spacecraft's SWE (Solar Wind Experiment) instrument\footnote{http://web.mit.edu/afs/athena/org/s/space/www/wind.html}, as shown in Fig. \ref{ACE}. It has been found (\citealt{Fujimoto2007}, \citealt{Yoshino2009}) that Suzaku spectra do not show strong SWCX signatures when the proton flux is below 4$\times$10$^{8}$ cm$^{-2}$ s$^{-1}$. As shown in Fig. \ref{ACE}, the proton flux only slightly exceeds 4$\times$10$^{8}$ cm$^{-2}$ s$^{-1}$ during the first observation (806084010), which is the closest to the cluster core, while the other observations are not affected. The high X-ray brightness of the ICM emission from this pointing, together with the fact that this is only a slight excess in the proton flux (the proton flux can increase by an order of magnitude above 4$\times$10$^{8}$ cm$^{-2}$ s$^{-1}$ during a strong flare), makes the effect of SWCX negligible. 

The mosaicked Suzaku image shown in Fig. \ref{annuliPKS} was obtained following the method described in \citet{Bautz2009}, adding together the images from the front illuminated detectors (XIS0, XIS3). Point sources which are bright enough to be resolved by the Suzaku pointings have been removed from this image and from the spectral extraction using 2.5 arcmin circular exclusion regions.

\subsection{Chandra data reduction}

The ACIS-I data were cleaned and calibrated using \textsc{CIAO-4.4}. The exposure corrected mosaicked image shown in Fig. \ref{annuliPKS} (right panel) was obtained using the \textsc{merge\_obs} script in \textsc{CIAO}. Point sources were identified using \textsc{wavdetect} using a range of wavelet radii between 1 and 16 pixels to ensure all point sources were detected.

\subsection{ROSAT data reduction}
\label{ROSATdatareduction}
The ROSAT PSPC observations were reduced using the Extended Source Analysis Software (ESAS, \citealt{Snowden1994}), and following the procedure described in \citet{Eckert2012}. The resulting exposure corrected mosaicked image for the R37 band (0.4-2.0keV) is shown in the right panel of Fig. \ref{annuliPKS}. When used later to derive the density profile the point sources were removed using the program \textsc{detect} with a constant threshold flux of 0.003 cts/s in the R37 band to ensure that the CXB is uniformly resolved across the whole field of view. In the mosaicked image in Fig. \ref{annuliPKS} the point sources are left in for display purposes only.

\section{Background Subtraction and Modelling}
\label{newbackgroundmodelling}

Accurate measurements of the ICM in the cluster outskirts requires an accurate knowledge of the components of the X-ray and non X-ray background, and an accurate understanding of how these are expected to vary spatially. Suzaku's low earth orbit gives it a low and stable non X-ray background which can be reproduced with 3 percent accuracy using night Earth observations \citep{Tawa2008}.  

\subsection{The CXB}
\label{CXB}

The cosmic X-ray background (CXB) consists of unresolved point sources, and is modelled as a powerlaw with index 1.4. As described in \citet{Bautz2009} and \citet{Walker2012_A2029}, our Suzaku data alone allow us to remove point sources down to the threshold detection limit of $S^{Suzaku}_{excl}$=10$^{-13}$ erg cm$^{-2}$ s$^{-1}$ in the 2-10 keV band, and these were removed during the spectral extraction using 2.5 arcmin circular exclusion regions. Following \citet{Walker2012_A2029}, we use the two-power-law model for the cumulative flux distribution of point sources found using ROSAT, Chandra and XMM-Newton observations in \citet{Moretti2003} to find the unresolved CXB flux remaining once the point sources resolved with Suzaku have been removed. Taking the total CXB flux in the 2-10 keV band to be 2.18 $\pm$0.13 $\times$ 10$^{-11}$ erg
cm$^{-2}$ s$^{-1}$ deg$^{-2}$ (from \citealt{Moretti2009} using \emph{Swift} data), and performing the integral,
\begin{eqnarray}
F_{\rm CXB} = 2.18 \pm 0.13 \times 10^{-11} - \int_{S_{\rm excl}}^{S_{\rm max}}
\Big(\frac{dN}{dS} \Big)
\times S ~ dS
\label{CXBlevelequation}
\end{eqnarray}
we find that following the removal of point sources down to the threshold flux $S^{Suzaku}_{excl}$ the remaining CXB level is 1.87 $\pm$ 0.13 $\times$10$^{-11}$ erg cm$^{-2}$ s$^{-1}$ deg$^{-2}$.

Outside $r$ = 50 arcmins, our Chandra observations allow point sources to be resolved down to a threshold flux of $S^{Chandra}_{excl}$=1.5$\times$10$^{-14}$ erg cm$^{-2}$ s$^{-1}$ in the 2-10 keV band across the whole ACIS-I field observed. This means that in the outskirts the unresolved CXB emission is reduced to 1.37 $\pm$ 0.08 $\times$10$^{-11}$ erg cm$^{-2}$ s$^{-1}$ deg$^{-2}$.

To account for the point sources resolved by Chandra in our spectral fitting to the Suzaku data, we used the ray-tracing simulator \textsc{xissim} to simulate the emission from the point sources resolved with Chandra in the Suzaku pointings for each detector. The spectra of the point sources extracted from the Chandra data are all in good agreement with a powerlaw of index 1.4, which was used as the input spectral model to \textsc{xissim}. We simulated exposures 1000 times the actual exposures to get good statistics. Spectra were then extracted from each simulated observation using the annuli shown in Fig. \ref{annuliPKS}, and the resolved point source contribution in each annulus was modelled using xspec, and included in the background model. This means that the only unresolved CXB emission outside $r=$50 arcmins is from point sources below 1.5$\times$10$^{-14}$ erg cm$^{-2}$ s$^{-1}$. 

Because the CXB consists of unresolved point sources, which are not uniformly distributed across the sky, the CXB level deviates from the mean when analysing small areas due to varying numbers of unresolved point sources. The expected deviation from the average value for a given observed solid angle ($\Omega$) resolved to a threshold flux $S_{excl}$ can be calculated using \citep{Bautz2009}, 
\begin{eqnarray}
\sigma^2_{\rm CXB} = (1/\Omega) \int_{0}^{S_{\rm excl}} \Big(\frac{dN}{dS}
\Big)
\times S^2 ~ dS
\label{CXBvarequation}
\end{eqnarray}
Our Chandra observations of the outskirts allow the CXB to be resolved to a threshold flux $\sim$ 10 times lower than that achievable with just Suzaku data. Integrating equation \ref{CXBvarequation}, we find that this reduces the uncertainty in the CXB level due to unresolved point sources by a factor of 2.3.   

Due to the large spatial extent of Centaurus, the extraction regions used in spectral analysis can be made large, which further reduces the expected variations in the CXB level for the regions analysed, (as well as significantly reducing the effects of PSF spreading between annuli). The expected 1 $\sigma$ variations in the 2-10keV CXB flux for the annuli investigated are shown in table \ref{CXB_fluctuations}. Later in section \ref{systematicerrors} we vary the CXB level of the background model through these ranges to calculate the systematic error of the CXB level uncertainty on the spectral fitting parameters.   

When performing the fits to the cluster emission in the outskirts (outside 50 arcmins) the low temperature of the cluster emission meant that the emission above 5 keV was purely from the CXB, which allowed us to check that the CXB level was consistent with the calculated value (see the spectral fits for the outermost 4 annuli in Fig. \ref{cluster_spectra}). In all cases the unresolved CXB level obtained by letting the CXB norm be a free parameter was highly consistent with the value calculated earlier (1.37 $\pm$ 0.08 $\times$ 10$^{-11}$ erg cm$^{-2}$ s$^{-1}$ deg$^{-2}$ for the regions outside 50 arcmins).

\begin{figure}
  \begin{center}
    \leavevmode
    \hbox{
      \epsfig{figure=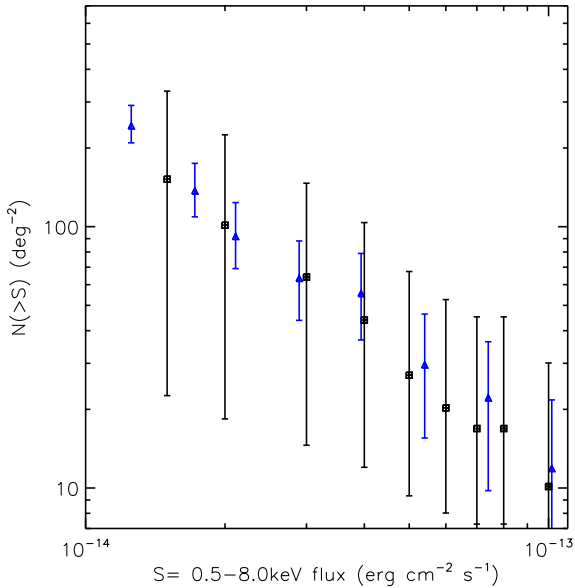, angle=0,
        width=\linewidth}
             
        }     
      \caption{Here we compare the cumulative source number counts obtained in our Chandra observations (black points) in the 0.5-8.0keV band with the same distribution obtained in the Chandra Deep Field South (CDFS) survey (blue points). We find agreement between the two, and there is no evidence for an excess in the number counts in our Chandra observations which would occur if some of the point sources were bright gas clumps rather than background AGN.  }
      \label{CDFS}
  \end{center}
\end{figure}

To test the possibility that some of the point sources identified with the \emph{Chandra} observations are actually bright gas clumps, we compared the cumulative number counts of point sources (logN-logS) with that of the Chandra Deep Field South (CDFS) survey field \citep{Lehmer2012}. Any excess over the source population in the CDFS survey would be evidence for additional bright sources, possibly due to gas clumping. The comparison is shown in Fig. \ref{CDFS}, where we compare the cumulative source number counts in the 0.5-8.0keV band observed in our Chandra fields (black points), with that observed in the CDFS (blue triangles). Poisson uncertainties in the total number of sources observed in our Chandra fields dominate the uncertainty, and the 1-$\sigma$ asymmetric confidence limits shown were calculated using the formulae presented in \citet{Gehrels1986} (their equation 9 was used for the upper limits, and their equation 14 was used for the lower limits). The source population is in complete agreement with the source population from the CDFS, and there is no evidence for an excess over the CDFS level which would occur if some of the point sources removed were bright gas clumps instead of background AGN.

\begin{table*}
  \begin{center}
  \caption{For each annulus we show the threshold flux for excluding point sources ($S_{excl}$), the unresolved CXB flux ($F_{CXB}^{unresolved}$), the resolved CXB flux ($F_{CXB}^{resolved}$), and the expected one sigma variations in the unresolved CXB flux each annulus ($\sigma_{unresolved}$) in the 2-10 keV band. }
  \label{CXB_fluctuations}

    \leavevmode
    \begin{tabular}{lllllllllllllll} \hline \hline

 Annulus &10-15 & 15-20 & 20-25 & 25-30 & 30-35 & 35-45  & 45-50 & 50-60 &60-70 &70-80 & 80-90 &90-105 \\ \hline
 $S_{excl}$ (10$^{-14}$ erg s$^{-1}$ cm$^{-2}$ deg$^{-2}$) & 10  &    10  &    10     & 10   &   10   &   10   &   10   &  1.5   &   1.5  &    1.5  &   1.5& 1.5 \\    
$F_{CXB}^{unresolved}$ (10$^{-12}$ erg s$^{-1}$ cm$^{-2}$ deg$^{-2}$) & 18.7  &  18.7  &   18.7  & 18.7 &  18.7   &   18.7   &   18.7   &  13.7   &  13.7  &   13.7  &   13.7& 13.7\\    
$F_{CXB}^{resolved}$ (10$^{-12}$ erg s$^{-1}$ cm$^{-2}$ deg$^{-2}$) & 0  &  0  &   0  & 0 &  0   &   0   &   0   &  2.7   &  3.6&2.4  &   7.5  &   3.3\\    
$\sigma_{unresolved}$ (10$^{-12}$ erg s$^{-1}$ cm$^{-2}$ deg$^{-2}$) & 3.8  &    3.3  &    3.4     & 3.4   &   3.3   &   2.9   &   3.4   &   1.1   &   1.1  &    1.1  &   1.2& 0.8\\    \hline

    \end{tabular}
  \end{center}
\end{table*}

\subsection{Modelling the soft foreground.}

\label{softforeground}

We use the Suzaku region outside 90 arcmin, together with RASS background region between 1.5 to 2.5 degrees\footnote{The RASS data were obtained from the X-ray background tool at http://heasarc.gsfc.nasa.gov/cgi-bin/Tools/xraybg/xraybg.pl} (corresponding to the region between 1.1-1.8$r_{200}$). We also use the ROSAT PSPC background regions around nearby NGC 4507. All of these background regions are well fit by a model of the form phabs*(powerlaw + apec(0.22keV)) + apec(0.12keV), which consists of an absorbed powerlaw to describe the CXB from unresolved point sources, an absorbed thermal component at 0.22keV to describe galactic halo (GH) emission, and an unabsorbed thermal component at 0.12keV to describe the Local Hot Bubble (LHB). Adding an additional absorbed thermal component, which is sometimes necessary to fit the data \citep{Snowden2008}, does not improve the quality of the fits. The temperatures of the galactic halo and local hot bubble were obtained by spectral fitting to the RASS data, and the metallicities of these components were fixed to 1 $Z_{\odot}$, while the redshifts of these components were fixed to zero. The different background regions studied are all found to be reasonably consistent and the best fit parameters shown in table \ref{GALvariations}.  We can use these to gain an understanding of the expected spatial variation of the soft foreground over the cluster, and we later calculate the systematic errors resulting from this uncertainty on the background model on the spectral fits in section \ref{systematicerrors}. When performing the background fits the normalisations of the LHB, GH and CXB components, and the temperatures of the LHB and GH components, were free parameters. 

\begin{table*}
  \begin{center}
  \caption{Soft foreground measurements from ROSAT PSPC and \emph{Suzaku} data. The units of the \textsc{apec} normalisations are $10^{-14}(4\pi)^{-1}D_{A}^{-2}(1 + z)^{-2} \int n_{e} n_{H} dV$, where   $D_{A}$ is the angular size distance (cm), and $n_{e}$ and $n_{H}$ are the electron and hydrogen densities (cm$^{-3}$) respectively, and these values are scaled for a circular area of sky of 20$'$ radius (1257 arcmin$^{2}$).}
  \label{GALvariations}

    \leavevmode
    \begin{tabular}{llllllllll} \hline \hline

       Position   &  0.22 keV \textsc{apec} norm (GH) & 0.12 keV \textsc{apec} norm (LHB)    \\ \hline

    SUZAKU N offset (90$'$-105$'$)  & 1.7$^{+0.5}_{-0.7}$ $\times$ 10$^{-3}$ & 1.5$^{+2.0}_{-0.5}$ $\times$ 10$^{-3}$ \\
    RASS 1.8-2.5 degree background  & 2.3$^{+0.4}_{-0.3}$ $\times$ 10$^{-3}$ & 1.0$^{+0.1}_{-0.1}$ $\times$ 10$^{-3}$ \\
    ROSAT NGC 4507                  & 1.9$^{+0.15}_{-0.15}$ $\times$ 10$^{-3}$ & 2.3$^{+0.3}_{-0.3}$ $\times$ 10$^{-3}$\\ \hline

    \end{tabular}
  \end{center}
\end{table*}

\subsection{Stray Light and PSF spreading}
\label{stray light}

To calculate the contributions in each annulus from neighbouring annuli due to PSF spreading and stray light, we simulated the contribution in each annulus from all of the other annuli using the background subtracted image of the cluster as input to the ray tracing simulator \textsc{xissim}, as was done in \citet{Walker2012_A2029} and \citet{Walker2012_PKS0745}. The results are tabulated in table \ref{psf_table}, where each row shows that contribution in each annulus from the annuli in the columns. The majority of the emission in each annulus originates from that annulus, and most of the external contribution is from the annulus immediately inside the annulus in question. 

Due to the roll angle of the observations, stray light from the core has been heavily reduced, such that in the outer annulus it contributes only 0.3 percent. When performing the spectral fits we tested the effect of including light spread from the core and from the adjacent annuli, and found this to have no effect on the best fitting temperatures, densities and metallicities. To perform this test we fitted all of the annuli simultaneously and modelled each annulus as consisting of emission from the annulus itself and all of the other annuli. Each annulus was modelled as the sum of the \textsc{apec} component originating from that annulus with those \textsc{apec} contributions from the other annuli, weighted by the fractions shown in table \ref{psf_table}.    

\begin{table*}
  \begin{center}
  \caption{Percentage contribution of flux in the rows' annulus from the columns' annulus due to PSF spreading and stray light.}
  \label{psf_table}
  
    \leavevmode
    \begin{tabular}{lllllllllllll} \hline \hline 
   & 0$'$-5$'$  & 	5$'$-10$'$	 & 10$'$-15$'$	 & 15$'$-20$'$ & 	20$'$-25$'$	 & 25$'$-30$'$  & 30$'$-35$'$  & 35$'$-45$'$ &  45$'$-50$'$ &  50$'$-60$'$ &  60$'$-70$'$ &  70$'$-80$'$ \\ \hline
0$'$-5$'$  &       \textbf{94} &   5.6 &  0.12 & 0.039 & 0.020 & 0.020 & 0.020 & 0.020 & 0.029 & 0.029 &0.0099 &0.0099  \\
 5$'$-10$'$	 &      12 &   \textbf{84} &   2.2 &  0.16 &  0.14 &  0.12 &  0.11 &  0.12 &  0.13 &  0.11 & 0.070 & 0.029   \\
10$'$-15$'$	 &      0.61 &   7.2 &   \textbf{80} &   9.9 &  0.34 & 0.099 & 0.099 &  0.14 &  0.12 &  0.16 &  0.13 & 0.089   \\
15$'$-20$'$ &      0.14 &  0.59 &   10.0 &   \textbf{80} &   7.3 &  0.41 &  0.22 &  0.23 &  0.14 &  0.14 &  0.11 & 0.079  \\ 
 20$'$-25$'$	 &     0.12 &  0.19 &  0.85 &   11.0 &   \textbf{80} &   4.7 &  0.59 &  0.47 &  0.38 &  0.25 &  0.14 &  0.12  \\ 
25$'$-30$'$  &      0.11 &  0.17 &  0.26 &  0.55 &   6.1 &   \textbf{80} &   10 &  0.47 &  0.29 &  0.29 &  0.28 &  0.20  \\
30$'$-35$'$  &      0.31 &  0.26 &  0.35 &  0.37 &  0.64 &   8.8 &   \textbf{82} &   5.4 &  0.28 &  0.26 &  0.19 &  0.17  \\
35$'$-45$'$ &      0.89 &  0.94 &  0.94 &  0.95 &   1.1 &   1.6 &   7.0 &   \textbf{80} &   3.2 &  0.66 &  0.44 &  0.41  \\
45$'$-50$'$ &      0.38 &  0.34 &  0.32 &  0.28 &  0.25 &  0.31 &  0.23 &   3.4 &   \textbf{87} &   6.5 &  0.28 &  0.26  \\
50$'$-60$'$ &       1.5 &   1.7 &   1.5 &   1.6 &   1.2 &   1.1 &   1.2 &  0.98 &   3.0 &   \textbf{83} &   1.2 &  0.61   \\
60$'$-70$'$ &      0.58 &  0.71 &  0.81 &  0.81 &  0.65 &  0.78 &  0.69 &  0.63 &  0.59 &   4.0 &   \textbf{84} &   4.8   \\
70$'$-80$'$ &      0.32 &  0.58 &  0.87 &   1.0 &   1.0 &   1.0 &   1.1 &   1.0 &  0.96 &  0.77 &   4.9 &   \textbf{86}   \\ \hline

    \end{tabular}
  \end{center}
\end{table*}

\section{Spectral Analysis}
\label{analysis}

Spectra were extracted in the annuli shown in Fig. \ref{annuliPKS} between radii 10-15, 15-20, 20-25, 25-30, 30-35, 35-40, 40-45, 45-50, 50-60, 60-70, 70-80 and 80-90 arcmins. These radii were chosen to ensure that each annulus contained at least 2000 counts following background subtraction. In all cases the annulus width is much larger than the PSF of Suzaku (HPD 2 arcmins) to prevent the effects of spillage between the annuli. The spectra from each detector (XIS0, XIS1, and XIS3) and each editing mode (3$\times$3 and 5$\times$5) were all fitted simultaneously for each annulus. 

We first performed projected fits for all of the annuli, modelling each annulus as an absorbed \textsc{apec} \citep{Smith2001} component. We fix the column density to the LAB survey \citep{LABsurvey} values for each pointing shown in table \ref{obsdetails}. As in \citet{Simionescu2011} and \citet{Urban2011}, we use the abundance tables of \citet{Grevesse1998}. 

\begin{figure}
  \begin{center}
    \leavevmode
    \hbox{
      \epsfig{figure=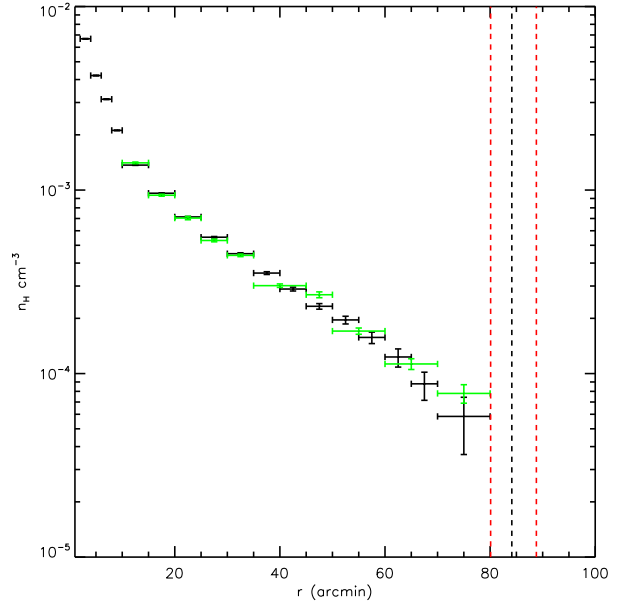, angle=0,
        width=\linewidth}
             
        }     
      \caption{Green shows Suzaku projected densities, black shows ROSAT projected densities (for full azimuthal coverage). The position of $r_{200}$ and its error is shown by the vertical dashed black and red lines respectively. }
      \label{compareROSATtoSuzaku}
  \end{center}
\end{figure}

The projected density profile, obtained from the \textsc{apec} normalisations during the spectral fitting, is shown in green in Fig. \ref{compareROSATtoSuzaku} and compared to the projected density profile obtained later in section \ref{comparetoROSAT} using the ROSAT PSPC data (which have higher azimuthal coverage) to ensure the Suzaku results are not biased by looking along one strip. We find strong agreement between the Suzaku and ROSAT projected density profiles out to 80 arcmins. We find no statistically significant cluster emission outside 80 arcmins.

Deprojected temperature and density profiles were obtained by modelling each annulus as the superposition of the ICM from the shell corresponding to that annulus with the emission projected onto the annulus from the shells exterior to it, with each component scaled according to the volume contributing to each annulus. The emission from each shell is modelled as an absorbed apec component. This emulates the xspec mixing model PROJCT, but allows for the use of background modelling and for the simultaneous fitting of data from multiple detectors (as in \citealt{Humphrey2011, Walker2012_A2029}). The deprojected densities are derived from the \textsc{apec} normalisations for each annulus obtained during the spectral fitting, using the volumes of the shells observed in each annulus. The spectral fits are shown in Fig. \ref{cluster_spectra}.

The deprojected temperature and density profiles are shown in the top two panels of Fig. \ref{T_and_d_profiles}. The temperatures found for the region within 0.03-0.2$r_{180}$ with XMM-Newton in \citet{Matsushita2011} are shown by the green points in the top panel of Fig. \ref{T_and_d_profiles}, showing excellent agreement with our Suzaku temperatures. The temperatures found for the NW sector using BeppoSAX in \citet{Molendi2002} are shown as the pink points, which again show excellent agreement. The red points show the temperatures in the NW sector obtained using Chandra data in \citet{Sanders2006}. 

Using our Suzaku data the metallicity can be constrained out to 50 arcmins. The metallicity profile, shown as the black points in the bottom panel in Fig. \ref{T_and_d_profiles}, falls from around 0.4$Z_{\odot}$ in the 10-15 arcmin annulus to 0.1$Z_{\odot}$ in the 45-50 arcmin annulus. Outside of this annulus the metallicity cannot be constrained and it is fixed to 0.1 Z$_{\odot}$ (shown by the blue lines). Therefore, despite the high central metallicity of the Centaurus cluster (reaching 3 Z$_{\odot}$ in the central 30 kpc, \citealt{Fabian2005}) the metallicity declines in the outskirts and is not higher than metallicity measurements for other clusters in the outskirts. For the Perseus cluster, \citet{Simionescu2011} found metallicities around 0.3$Z_{\odot}$ in the outskirts, while for the less massive Virgo cluster \citet{Urban2011} found a lower outskirts metallicity of 0.1$Z_{\odot}$. This may suggest that less massive clusters have lower metallicities in the outskirts. For instance in \citet{Sun2012} it has been found by comparing the abundance profiles of samples of galaxy clusters and galaxy groups (their Fig. 4) that galaxy groups are iron poorer than galaxy clusters in the region 0.3-0.7$r_{500}$.      

The metallicities found in the region within 0.03-0.2$r_{180}$ with XMM-Newton in \citet{Matsushita2011} are shown by the green points in the bottom panel of Fig. \ref{T_and_d_profiles} and agree with the Suzaku metallities in this region. The metallicities found in the NW sector in \citet{Sanders2006} are shown as the red points. The pink points show the metallicities obtained for the NW sector using BeppoSAX data in \citet{Molendi2002}, but we note that this study used the abundance tables of \citet{Anders1989}, which yield Fe abundances lower by a factor of around 1.4 than the abundance tables of \citet{Grevesse1998} which we use.  
 
\subsection{Entropy and pressure profiles} 
 \label{entropyandpressureforCentaurus}
 
\begin{figure}
  \begin{center}
    \leavevmode
    \hbox{
      \epsfig{figure=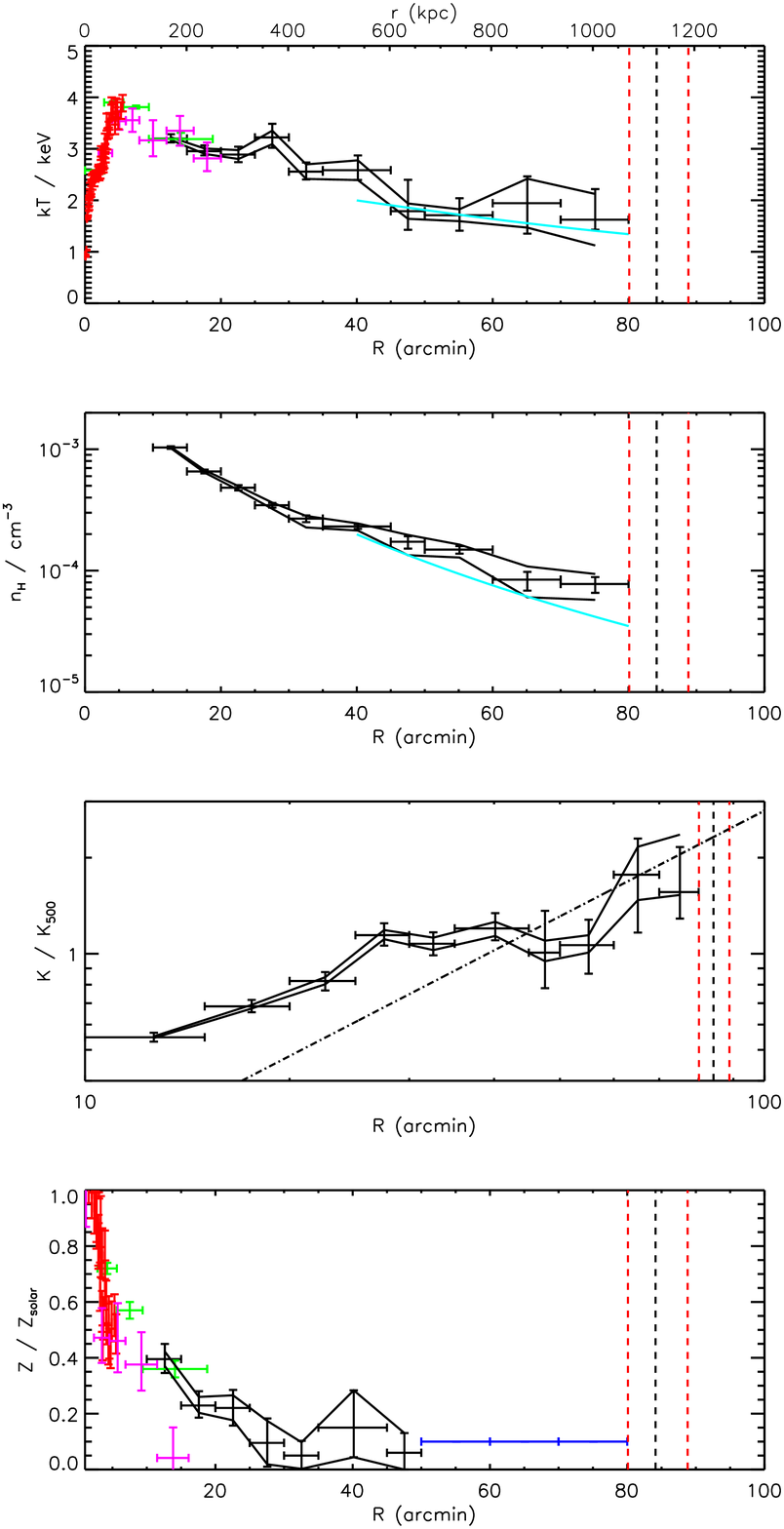, angle=0,
        width=\linewidth}
        }    
      \caption{\emph{First panel}: Deprojected temperature profile. The green points are the temperatures from \citet{Matsushita2011} obtained with XMM-Newton, the pink points are the temperatures obtained for the NW sector in \citet{Molendi2002} using BeppoSAX, and the red points are the temperatures obtained in the NW sector with Chandra in \citet{Sanders2006}. The cyan lines shows the temperatures and densities predicted by combining the baseline entropy profile of \citet{Voit2005} and the universal pressure profile of \citet{Arnaud2010}. \emph{Second panel}:  Deprojected density profile, \emph{Third panel}: Deprojected entropy profile. The dashed line shows the $r^{1.1}$ powerlaw relation predicted for purely gravitational hierarchical structure formation (from \citet{Voit2005}), scaled using the self similar entropy, $K_{500}$ from equation \ref{eqn:K500}. \emph{Fourth panel}: Metallicity profile. Green points are from \citet{Matsushita2011} using XMM-Newton; pink points are for the NW sector using BeppoSAX from \citet{Molendi2002}; the red points are for the NW using Chandra from \citet{Sanders2006} . The blue lines show where the metallicity has been fixed in the outskirts to 0.1 Z$_{\odot}$. In all of the panels the solid black lines show the systematic errors calculated in section \ref{systematicerrors}. The dashed vertical black line shows the value of $r_{200}$ calculated in section \ref{massanalysis}, with the vertical dashed red lines showing the error in $r_{200}$.   }
      \label{T_and_d_profiles}
  \end{center}
\end{figure}

The deprojected entropy ($K=kT/n_{e}^{2/3}$) is calculated from the deprojected temperature and density profiles. The deprojected entropy profile is shown in Fig. \ref{T_and_d_profiles}, where it is compared to the baseline entropy profile for purely gravitational collapse from \citet{Voit2005},
\begin{equation}
K(R)/K_{500} = 1.47(r/r_{500})^{1.1}.
\label{eqn:KR}
\end{equation}
 Here we have accounted for the hydrostatic mass estimates being biased low by $\sim$ 13 $\pm$ 16 percent on average (because these neglect the non-thermal pressure support in the ICM), by increasing the factor in equation \ref{eqn:KR} from 1.42 to 1.47 as described in \citet{Pratt2010}. The baseline profile has been scaled using the self-similar entropy at $r_{500}$ (equation \ref{eqn:K500}) as in \citet{Pratt2010},
\begin{eqnarray}
\label{eqn:K500}
K_{500} &=& 106 \ {\rm keV\ cm}^{-2} \left( \frac{M_{500}}{10^{14}\,h_{70}^{-1}\,M_\odot} \right)^{2/3}\, \left(\frac{1}{f_b}\right)^{2/3}\, \nonumber \\ 
&& \times E(z)^{-2/3}\, h_{70}^{-4/3}
\end{eqnarray}
using $f_{b}$=0.15 as in \citet{Pratt2010}, and using the $M_{500}$ value of 1.2$\times$10$^{14}$ $M_{\odot}$ which we find later in our mass analysis in section \ref{massanalysis}. 

  We find that the entropy exceeds the baseline level within 40 arcmins, and flattens towards it at $\sim$0.5$r_{200}$, which is similar to the behaviour found for the REXCESS clusters in \citet{Pratt2010}. Outside 50 arcmins ($\sim$ $r_{500}$), the entropy is systematically slightly below the baseline relation, possibly the results of gas clumping, though in the outermost two annuli it is in good agreement with the baseline relation. Later, in section \ref{UEP}, we will compare the entropy profiles of other clusters explored in the outskirts with Suzaku when scaled using self-similar scaling relations.

\begin{figure}
  \begin{center}
    \leavevmode
\hbox{
      \epsfig{figure=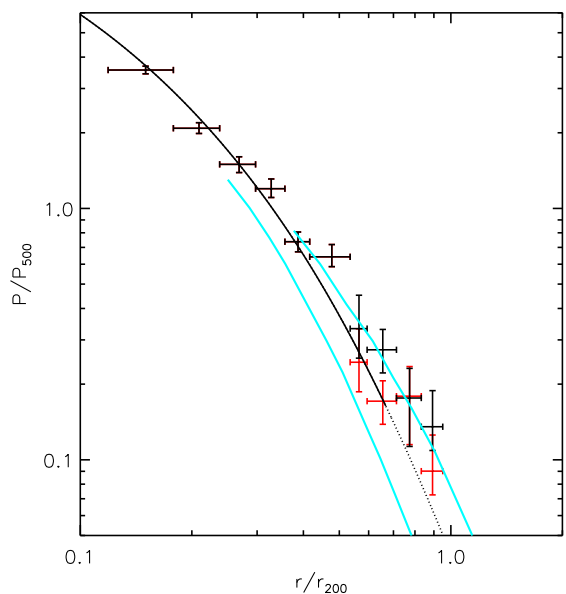, angle=0,
        width=\linewidth}
      }
        
      \caption{Deprojected pressure profile, which is consistent with the universal pressure profile found in \citet{Arnaud2010}, shown as the solid black line inside $r_{500}$ (where it is based on observations), and as a dotted line outside $r_{500}$ (where it is based on simulations). The uncertainties on the universal pressure profile in the outskirts shown in \citet{Lapi2012} are shown by the cyan lines. The red points show the effect of correcting the pressures by the clumping values needed to make the entropy profile agree with the baseline profile in the outskirts. This clumping correction lowers the pressures and brings them into even better agreement with the universal pressure profile.}
      \label{pressureprofile}
  \end{center}
\end{figure}

The deprojected pressure profile ($P=kn_{e}T$) is shown in Fig. \ref{pressureprofile}, and is compared with the universal pressure profile of \citet{Arnaud2010},
\begin{align}
\label{eq:puniv}
P(r)= P_{500}\left[\frac{M_{500}}{3 \times 10^{14}\,{\rm h^{-1}_{70}}\,M_{\odot}}\right]^{ \alpha_{\rm P}+\alpha^{\prime}_{\rm P}(x)} \mathbb{P}(x)
\end{align}
where,
\begin{align}
\mathbb{P}(x) &= \frac{P_{0} } { (c_{500}x)^{\gamma}[1+(c_{500}x)^\alpha]^{(\beta-\gamma)/\alpha} } \\ 
(P_{0} ,c_{500},\gamma,\alpha,\beta) &=  (8.403\,{\rm h_{70}^{-3/2}},1.177,0.3081,1.0510,5.4905) \\
\alpha_{\rm P}+\alpha_{\rm P}^{\prime}(x)  &=  (0.10 + \alpha_{\rm P})\left[1- \frac{ (x/0.5)^{3  }}{ 1.+(x/0.5)^{3}}\right] \\
x &= r/r_{500}
\label{eq:pgnfw}
\end{align}
after being appropriately scaled by the characteristic pressure, P$_{500}$,
\begin{align}
P_{500}  =  1.65\times10^{-3}\,h(z)^{8/3}\,\left[\frac{\mathrm{M_{500}}}{3\times10^{14}\,{\rm h^{-1}_{70}}\,\rm M_{\odot}}\right]^{2/3}~~{\rm h_{70}^{2}\,\keV\,cm^{-3}}
\label{eq:p500}
\end{align}
The $(M_{500} / 3 \times 10^{14}\rm h^{-1}_{70} M_{\odot})^{ \alpha_{\rm P}+\alpha^{\prime}_{\rm P}(x)}$ term describes the deviation from the self similar scaling relation for pressure. From equation (8) we see that when $x=r/r_{500}>1$ the index $\alpha_{\rm P}+\alpha_{\rm P}^{\prime}(x)$ tends to zero, so outside $r_{500}$ this term is negligible. 

Whilst the observed pressure profile agrees with the universal pressure profile within the spread of the simulations, it is systematically slightly higher than the universal pressure profile in the outskirts, (as has also been found for the Perseus cluster in \citealt{Simionescu2011}, for the fossil group RX J1159+5531 in \citealt{Humphrey2011} and for PKS 0745-191 in \citealt{Walker2012_PKS0745}), which further suggests that gas clumping may be having an effect on our measured profiles. By causing the gas density to be overestimated, gas clumping would cause the gas pressure to be overestimated. 
  
We can combine the baseline entropy profile of \citet{Voit2005} and the universal pressure profile of \cite{Arnaud2010}, appropriately scaled by the self similar entropy ($K_{500}$) and pressure ($P_{500}$) respectively, to produce predictions for the temperature and density profiles in the outskirts as follows, 
\begin{eqnarray}
\label{ArnaudVoit_Tandd}
kT(r) &=& P(r)^{2/5} K(r)^{3/5} \\
n_{H}(r) &=& (1/1.2) P(r)^{3/5} K(r)^{-3/5}
\end{eqnarray}
and these predictions are shown as the cyan lines in the temperature and density plots in Fig. \ref{T_and_d_profiles}. We see that the temperatures agree with this prediction, but the densities are higher than it. This would appear to support the suggestion that it is the overestimate of the gas density, possibly due to gas clumping, that causes the entropies to be slightly lower than the baseline level and the pressures to be slightly higher than the universal pressure profile in the outskirts. 

In Fig. \ref{pressureprofile} we show that if the pressures are corrected by the clumping factor needed to make the entropy profile agree with the baseline profile in the outskirts (shown later in Fig. \ref{clumping}), then they agree better with the universal pressure profile. These clumping factors have been calculated using the assumption that the cause of the entropy profile lying below the baseline entropy profile in the outskirts is purely due to the overestimate of the gas density due to clumping. Defining the clumping factor as $C= \frac{\langle n_{gas}^{2} \rangle}{\langle n_{gas}  \rangle^{2}  }$ \citep{Nagai2011}, the observed density is overestimated by a factor of $\sqrt{C}$. If the true entropy in the outskirts is given by the baseline entropy profile (equation \ref{eqn:KR}) then this means that using the observed temperatures and densities at radii R ($kT(R)$ and $n_{e}(R)$) we can determine C(R) using;
\begin{eqnarray}
K(R)/K_{500} = 1.47(R/r_{500})^{1.1} = \frac{kT(R)}{K_{500}(n_{e}(R)/\sqrt{C(R)})^{2/3}}  
\label{clumpingcalc}         
\end{eqnarray}
\begin{eqnarray}
\sqrt{C(R)}=n_{e}(R) \left(1.47(R/r_{500})^{1.1}   \frac{K_{500}}{kT(R)}          \right)^{3/2}
\label{clumpingeqn}         
\end{eqnarray}

If gas clumping is present and the gas clumps are in pressure equilibrium with the surrounding ICM, then they would be expected to be colder than the surrounding ICM, and therefore bias the temperature low (this is also discussed in \citealt{Urban2011}). However we have found that for Centaurus the temperature profile in the outskirts is in good agreement with the temperature profile obtained by assuming the universal pressure profile and the baseline entropy profile, and that the deviations from the universal pressure profile and the baseline entropy profile appear to be purely the result of a density excess. This may indicate that the gas clumps are not in pressure equilibrium with the surrounding ICM. For instance it is likely that any gas clumps will be falling into the cluster and moving through the ICM, so ram pressure will help to support the clumps, and this may be the dominant support mechanism.  

\subsection{Comparing to ROSAT data}
\label{comparetoROSAT}
Following the approach of \citet{Eckert2012} we convert the surface brightness profile from the ROSAT PSPC mosaic into a projected density profile. As described in section \ref{ROSATdatareduction}, point sources were removed using the program DETECT with a constant threshold count rate of 0.003 cts/s in the R37 band to ensure that the CXB is resolved uniformly across the whole field of view, as in \citet{Eckert2012}. The background regions of the offset pointing of NGC 4507 shown in Fig. \ref{annuliPKS} were used to obtain the background level. The projected density profile was obtained using the ROSAT PSPC response and the Suzaku temperatures for each annulus to convert the count rates in each annulus into an apec normalisation. 

As the Suzaku observations have small azimuthal coverage, it is important to understand whether they are representative of the cluster as a whole. We find that the projected density profiles are highly consistent, as shown in Fig. \ref{compareROSATtoSuzaku}.

\subsection{Systematic errors}
\label{systematicerrors}

We need to calculate the systematic errors on the deprojected temperature, density and entropy profiles resulting from the uncertainty of the X-ray background parameters (the GH, LHB and CXB normalisations) which we have quantified in section \ref{newbackgroundmodelling}, and the uncertainty in the NXB level, which from \citet{Tawa2008} is $\pm$ 3 percent. To achieve this we produce 10000 realisations of the background model and the NXB level (as done in \citealt{Walker2012_A2029} and \citealt{Walker2012_PKS0745}), allowing all of the background parameters to vary simultaneously within their variances calculated earlier (and taking into account the covariance between background model parameters), and perform the deprojection in xspec for each realisation. This allows the uncertainty on the background model to be folded through the deprojection, resulting in a complete propagation of the errors. This allows a more realistic estimate of the systematic error to be obtained than is achieved by varying only one background parameter and leaving the others fixed to their best fit values. 

From the distribution of the 10000 profiles we can find the 1 $\sigma$ systematic errors, and these are shown by the solid lines in Fig. \ref{T_and_d_profiles}. These systematic errors also include the effect of varying the contamination on the optical blocking filter by $\pm$10 percent, which was estimated by using the ftool \textsc{xiscontamicalc} to modify the ARFs (as in \citealt{Akamatsu2011}), and was found to be a negligible effect much smaller than the statistical errors. The systematic errors also include the effect of varying the column density by $\pm$ 20 percent (the range of the values from the LAB survey for the observations as shown in table \ref{obsdetails}), and of varying the metallicity in the range 0.0-0.3$Z_{\odot}$ in the outer three annuli when performing the fits.  In all cases the total systematic errors are less than or equal to the statistical errors.  

\subsection{Large scale motions}

To search for the possibility of large scale gas motions, we divided the ROSAT image by the azimuthal average surface brightness profile (as done in \citealt{Simionescu2012} for the Perseus cluster). We find no significant features in the outer regions. In the central regions, the only significant features we find are those which have already been studied in \citet{Churazov1999} (their Fig. 3, right panel), and which our direction of study has avoided. \citet{Churazov1999} found an enhancement around 20 arcmins to the south east of the core, and that this enhancement corresponds to one of the subgroups in Centaurus (Cen 45, which contains the second brightnest galaxy, NGC 4709). The ASCA temperature map of the central regions also indicated that the enhancement in surface brightness coincides with an enhancement in temperature to the south east. \citet{Churazov1999} concluded that this higher temperature arose due to the heating caused by the interaction between the Cen 45 subgroup as it merges with the main cluster. This merging activity may be responsible for the cold front to the west of the core studied in depth with Chandra in \citet{Fabian2005}, by producing a sloshing motion of the core. 

\citet{Takahashi2009} and \citet{Lovisari2011} produced temperature maps of the central regions of Centaurus with full azimuthal coverage using XMM-Newton data and found that the north western sector has a slightly lower temperature than the other directions, however these observations only extend out to 12 arcmins (160 kpc$=$0.14$r_{200}$). This was also found in the Chandra temperature map of \citet{Lagana2010}, but this only extends out to 50 kpc (0.05$r_{200}$). Since the surface brightness is only weakly dependent on the temperature, any azimuthal variations in temperature will have a negligible effect on the observed gas mass.   

\section{Mass Analysis}
\label{massanalysis}

Assuming the ICM to be in hydrostatic equilibrium, and assuming the cluster is spherically symmetric, the total mass within radius, $r$, is given by \citep{Vikhlinin2006}

\begin{equation}
\label{hydroeq}
M(<r)=-\mathrm{3.68}\times\mathrm{10}^{\mathrm{13}}M_{\odot}T(r)r\left(\frac{d\: \mathrm{ln} \:n_{\rm
    H}}{d \:\mathrm{ln} \:r} + \frac{d\:\mathrm{ln}\:T}{d\:\mathrm{ln}\:r}\right) \\
\end{equation}

We assume the total mass is described by an NFW profile \citep{NFW1997}, which is well motivated by numerical simulations of gravitational collapse, 

\begin{eqnarray}
\rho(r)&=&\frac{\rho_0}{r/r_{\rm s}\left(1+(r/r_{\rm s})\right)^2} \\
\rho_0 &=& 200\rho_c c_{200}^{3}/3(\mathrm{ln}(1+c_{200}) - c_{200}/(1+ c_{200})) 
\end{eqnarray}

We follow the same mass analysis method as \citet{Walker2012_PKS0745}, which is based on that used in \citet{Schmidt2007}. We use the deprojected density profiles to predict the temperature profile assuming hydrostatic equilibrium and the NFW profile for the total mass, moving inwards from the outermost annulus (the results are unchanged if the method starts from the innermost annulus and moves outwards). We use the XMM-Newton temperatures near the core to better resolve the central regions.

The best fitting mass profile produces the best fitting temperature profile which minimises the $\chi^{2}$ statistic

\begin{eqnarray}
\label{chisqequation}
\chi^{2} = \sum_{i}\dfrac{(T_{\mathrm{calculated},i} -T_{\mathrm{actual},i})^2}{\sigma_{T_{\mathrm{actual},i}}^{2}}
\end{eqnarray}

As described in \citet{Allen2008}, because this method does not involve the use of parametric fitting functions for the temperature and gas density profiles, it avoids the strong priors that these place on the mass determination and which complicate the interpretation of results (and can lead to an underestimate of uncertainties). We stress that there is no robust prior expectation for the form the temperature and density profiles in the outskirts of clusters, and so the non-parametric approach we employ is important to ensure the results are not severely restricted by priors. 

To propagate the uncertainty of the density profile into the measurement of the gas mass fraction and the NFW parameters, we follow \citet{Walker2012_PKS0745} and repeat the mass analysis 10000 times using different realisations of the density profile distributed by the combined statistical and systematic errors shown in Fig. \ref{T_and_d_profiles}. For each iteration the cumulative gas mass fraction is calculated and the NFW best fitting parameters. The best fit values, together with the 1 $\sigma$ errors were then calculated from the resulting distributions. We find that the NFW profile describes the total mass profile well. We find $c_{200}$=5.9$^{+1.8}_{-1.4}$, $r_{s}$=191$^{+70}_{-52}$kpc, $r_{200}=1130^{+62.5}_{-54}$kpc$=84^{+4.6}_{-4.0}$arcmins and $M_{200}=1.6^{+0.3}_{-0.2} \times 10^{14}M_{\odot}$, and the fitting statistic is $\chi^{2}$/d.o.f = 12.5/12.  

To compare the calculated mass value with the M-T scaling relation of \citet{Arnaud2005}, we found the spectroscopic temperature in the region $0.1<r<0.5r_{200}$, which is $kT(0.1<r<0.5r_{200})=$ 3.0$_{-0.1}^{+0.1}$ keV, (these errors include the systematic errors in the background modelling). Using the scaling relation $M_{200}/10^{14}\mathrm{M_{\odot}}=5.34 \pm 0.22 \times (kT/5 \mathrm{keV})^{1.72 \pm 0.10}/h(z)$ we find $M_{200}^{Arnaud}$=2.2$^{+0.2}_{-0.2} \times 10^{14}$M$_{\odot}$, which is slightly higher but in reasonable agreement with our calculated mass given the scatter around the best fit relation for the M$_{200}$-T relation in \citet{Arnaud2005}. 

\begin{figure}
  \begin{center}
    \leavevmode
\hbox{
      \epsfig{figure=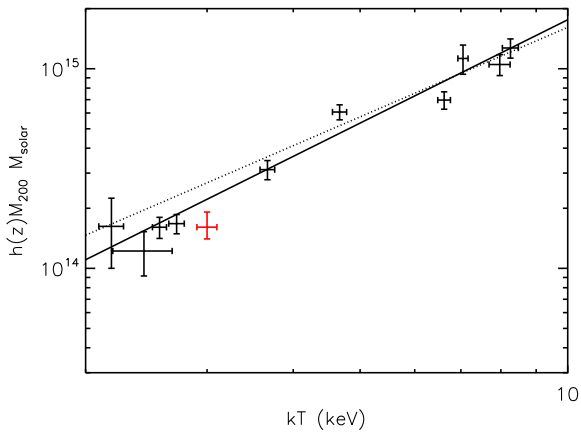, angle=0,
        width=\linewidth}
         }
        
      \caption{The $M_{200}-T$ plot from \citet{Arnaud2005}, with our value for the Centaurus cluster added as the red point. The solid line shows the best fit powerlaw relation in \citet{Arnaud2005} for all of the clusters, while the dashed line is the best fit powerlaw relation in \citet{Arnaud2005} for the clusters with $kT(0.1<r<0.5r_{200})$ $>$ 3.5keV. }
      \label{Arnaud2005}
  \end{center}
\end{figure}

Because the clusters studied in \citet{Arnaud2005} were mostly only studied out to a maximum radius of $\sim$0.6 $r_{200}$, the $M_{200}$ values used to obtain the $M_{200}-T$ scaling relation were derived by extrapolating out the best fitting NFW models to $r_{200}$. As described in \citet{Arnaud2005}, the quality of the powerlaw best fit decreases during this extrapolation, with the null hypothesis probability for the whole sample decreasing from 0.32 at $\delta$=2500 to 0.07 at $\delta$=200. In Fig. \ref{Arnaud2005} we plot Centaurus on the $M_{200}-T$ plot from \citet{Arnaud2005} as the red point, and we see that it is in reasonable agreement given the scatter around the best fit relation. 

\begin{figure}
  \begin{center}
    \leavevmode
\hbox{
      \epsfig{figure=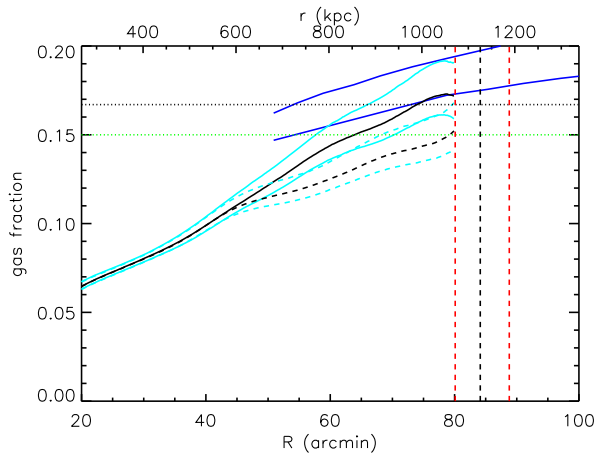, angle=0,
        width=\linewidth}
         }
        
      \caption{The cumulative gas mass fraction profile is shown as the solid black line with its error shown by the solid cyan lines. The dashed black and cyan lines show the effect of performing a clumping correction to make the entropy profile in the outskirts agree with the baseline profile. The horizontal dashed black line shows the mean cosmic baryon fraction of 0.167 found using WMAP data in \citet{Komatsu2011}, while the horizontal dashed green line shows the expected hot gas fraction when we account for 12 percent of baryons being in stars. The vertical dashed black line shows $r_{200}$, and its error is in red. The blue lines are the upper and lower limits of the gas mass fraction outside 0.6$r_{200}$ found using Planck in \citet{PlanckV2012} for a stacked sample of 62 clusters (using the \citealt{Vikhlinin2006} temperature profile), which agrees with the Centaurus profile in the outskirts. }
      \label{gasmassprofile}
  \end{center}
\end{figure}

\begin{figure}
  \begin{center}
    \leavevmode
\hbox{
      \epsfig{figure=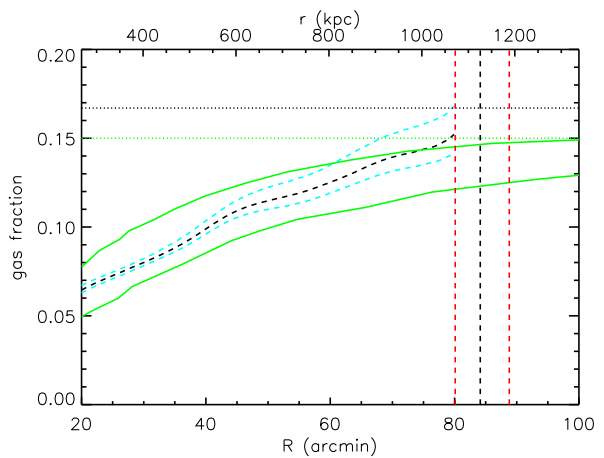, angle=0,
        width=\linewidth}
         }
        
      \caption{The same as Fig. \ref{gasmassprofile} but here we compare the clumping corrected gas mass fraction profile (dashed black line with errors in cyan) with the expected range in $f_{gas}$ obtained in the simulations of \citet{Young2011} for clusters with an average T in the 2.5-5.0keV range, shown by the green lines.  }
      \label{gasmassprofile_Young}
  \end{center}
\end{figure}

The gas mass fraction profile is shown in Fig. \ref{gasmassprofile}, and is found to rise to the mean cosmic baryon fraction at the highest radius observed. As shown in Fig. \ref{gasmassprofile}, the gas mass fraction in the outskirts agrees with that found using Planck data for a stacked sample of 62 clusters in \citet{PlanckV2012} (when the \citealt{Vikhlinin2006} temperature profile is assumed when obtaining the density profiles from the Planck data). In Fig. \ref{gasmassprofile} we show the effect of correcting the gas mass fraction by the clumping factors calculated for Centaurus in section \ref{UEP}, which are required to make the entropy profile agree with the baseline entropy profile in the outskirts. We see that performing such a correction reduces the gas mass fraction so that it agrees with the expected hot gas fraction (which takes into account that 12 percent of the baryons should be in stars). In Fig \ref{gasmassprofile_Young} we see that the clumping corrected gas mass fraction agrees well with the predictions of \citet{Young2011} for clusters in the temperature range 2.5-5.0 keV. 

\section{Universal Entropy Profile}
\label{UEP}
In \citet{Walker2012_UEP} we reported that the entropy profiles of clusters explored to $r_{200}$ with Suzaku, XMM-Newton and Chandra have the same shape, flattening off from a powerlaw relation above $\sim$0.5$r_{200}$. In \citet{Walker2012_UEP} the entropy profiles were scaled by their value at 0.3$r_{200}$, where all of the profiles demonstrated a powerlaw increase. Greater insight can however be gained by scaling the entropy profiles by the self-similar entropy at $r_{500}$ ($K_{500}$), as this allows both the shape and normalisation of the entropy profiles to be compared with theoretical expectations. Similar flattening has for instance been observed in the REXCESS sample of clusters obtained with XMM-Newton (\citealt{Pratt2010}), however the entropy is enhanced within $r_{500}$, and the flattening only acts to bring the entropy back into agreement with the baseline entropy profile of \citet{Voit2005} at $r_{500}$. 

\begin{table*}
  \begin{center}
  \caption{Sample of galaxy cluster outskirts observations used. Masses marked with an asterisk were calculated using the $M_{500}-T$ scaling relation of \citet{Arnaud2005}. Masses marked with $^{\dagger}$ were taken from the values measured in \citet{Arnaud2005} for those clusters.}
  \label{Cluster_sample}
  
    \leavevmode
    \begin{tabular}{lllllll} \hline \hline
    Cluster &z& Reference & Plot symbol& $M_{500}$/10$^{14}$ $M_{\odot}$\\ \hline
    Abell 1689 & 0.183& \citet{Kawaharada2010} &Red square&11.4\\
    Abell 2029 &0.0767 & \citet{Walker2012_A2029} &Red square&7.2\\ 
    Abell 2142 & 0.0899& \citet{Akamatsu2011} &Blue square&8.0\\
    Hydra A &0.0539 & \citet{Sato2012} &Cyan square&1.5\\
    Perseus & 0.0183& \citet{Simionescu2011} &Pink square&4.8\\
    PKS 0745-191 & 0.1028& \citet{Walker2012_PKS0745} &Grey square&7.3\\
    Abell 1835 &0.253 & \citet{Bonamente2012} & Black square&7.8\\
    Abell 2204 &0.152 & \citet{Sanders2009} & Black triangle&8.39$^{\dagger}$\\
    Abell 1795 &0.063 & \citet{Bautz2009} &Red triangle &4.1\\
    Virgo & 16.1 Mpc & \citet{Urban2011} & Green crosses&1.02*\\
    Abell 1413 & 0.143 & \citet{Hoshino2010} &Blue triangle&4.8$^{\dagger}$\\
    Centaurus &0.0109 & This work &Black crosses&1.2\\ 
    RX J1159+5531 &0.081 & \citet{Humphrey2011} &Pink triangles&0.63\\ 

      \hline

    \end{tabular}
  \end{center}
\end{table*}

To scale the entropy profiles we follow the approach of \citet{Pratt2010}, and scale the entropy profiles by the self-similar entropy at $r_{500}$ (equation \ref{eqn:K500}), and we compare this to the baseline entropy profile assuming only gravitational physics obtained in \citet{Voit2005} (equation \ref{eqn:KR}). The clusters used are tabulated in table \ref{Cluster_sample}. The $M_{500}$ values are taken from the values quoted in the papers listed in table \ref{Cluster_sample}, however for the Virgo cluster \citep{Urban2011} no mass analysis had been performed, so $M_{500}$ is calculated using the scaling relation of \citet{Arnaud2005} and the mean temperature of 2.3keV reported in \citet{Urban2011}. 

\begin{figure}
  \begin{center}
    \leavevmode
 
       \epsfig{figure=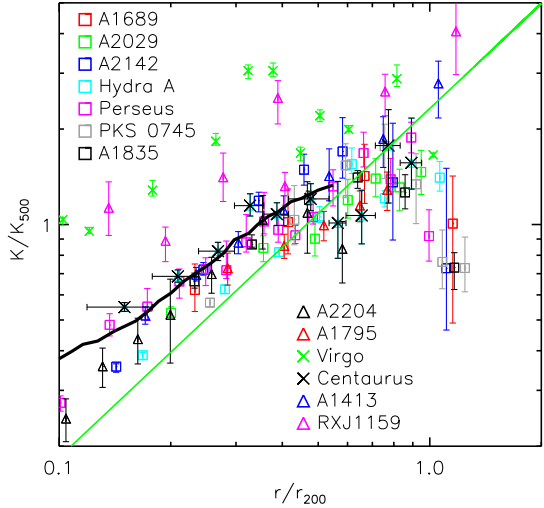, angle=0, width=\linewidth}
        
      \caption{Entropy profiles of clusters explored with Suzaku, XMM-Newton and Chandra in the outskirts, scaled by the entropy at $r_{500}$ predicted by self-similar scaling relations. The solid green line shows the baseline entropy profile from \citet{Voit2005} (equation \ref{eqn:KR}) calculated using only gravitational physics. The black line shows the median entropy profile from the REXCESS cluster sample in \citet{Pratt2010}.} 
      \label{entropy_compare}
  \end{center}
\end{figure}

\begin{figure}
  \begin{center}
    \leavevmode
 
       \epsfig{figure=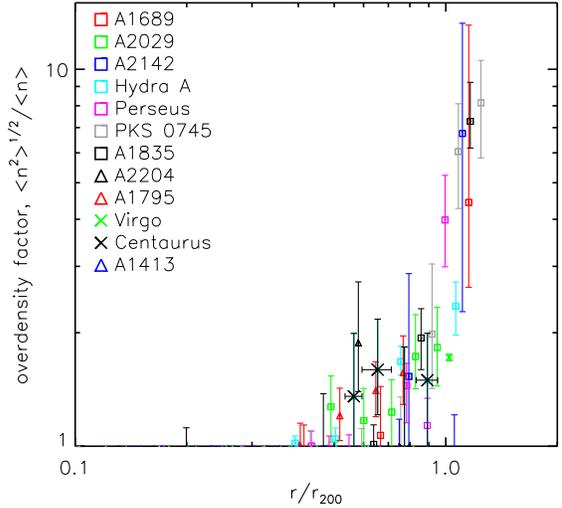, angle=0, width=\linewidth}
        
      \caption{Profile of the factors by which the gas density would need to be overestimated if gas clumping is the sole cause of the measured entropy profiles lying below the baseline entropy profile in the outskirts. } 
      \label{clumping}
  \end{center}
\end{figure}

In Fig. \ref{entropy_compare}, we find that the entropy within $r_{500}=0.659r_{200}$\footnote{The relation $r_{500}=0.659r_{200}$ assumes an NFW profile and the mean concentration parameter from the sample of \citet{Pointecouteau2005}, as used in \citet{Pratt2010}} is in excess of the baseline level for most of the clusters, and in general agrees with the baseline level at $r_{500}$. The behaviour inside $r_{500}$ roughly agrees with the median profile from \citet{Pratt2010}, shown by the solid black line, though we note that the REXCESS sample contained many lower mass systems which have acted to increase the median entropy level compared to our sample of mostly massive clusters. The low mass group 
RX J1159+5531 and the Virgo cluster demonstrate the largest entropy excess. We note that the Virgo entropies are in reasonable agreement with the baseline entropy profile in the outskirts (though the scatter for the Virgo datapoints is large). 

However outside $r_{500}$ the entropies are systematically below the baseline prediction using only gravitational physics. One possibility is that gas clumping (which causes the gas density to be overestimated) is causing the entropies outside $r_{500}$ to lie below the baseline entropy profile. In Fig. \ref{clumping} we show the factor by which the gas density needs to be overestimated in order for the entropy to agree with the baseline profile using equation \ref{clumpingeqn} for each cluster.  

\begin{figure}
  \begin{center}
    \leavevmode
 
       \epsfig{figure=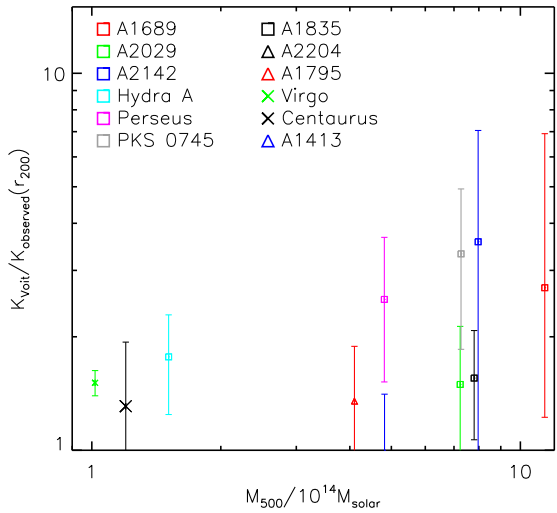, angle=0, width=\linewidth}
        
      \caption{Plotting the decrement of the entropy below the Voit baseline entropy profile at $r_{200}$ against cluster mass. There is no statistically significant correlation. } 
      \label{clumpingmassdependence}
  \end{center}
\end{figure}

The simulations of \citet{Nagai2011} found the predicted clumping level to be mildly mass dependent, with the more massive clusters having slightly higher clumping factors (thus causing a greater underestimate of the entropy). To search for signs of this we plot in Fig. \ref{clumpingmassdependence} the mass dependence of the entropy decrement below the baseline level ($K_{Voit}/K_{observed}$) near $r_{200}$ for the clusters studied. Due to the large errors and scatter, and the small sample size, there is no statistically significant correlation between the entropy decrement below the baseline level and the cluster mass.

\begin{figure}
  \begin{center}
    \leavevmode
 
\vbox{
      \epsfig{figure=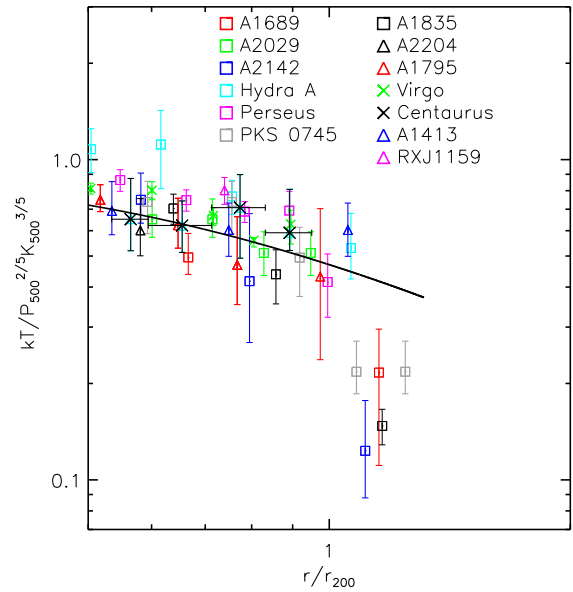, angle=0,
        width=\linewidth}
              \epsfig{figure=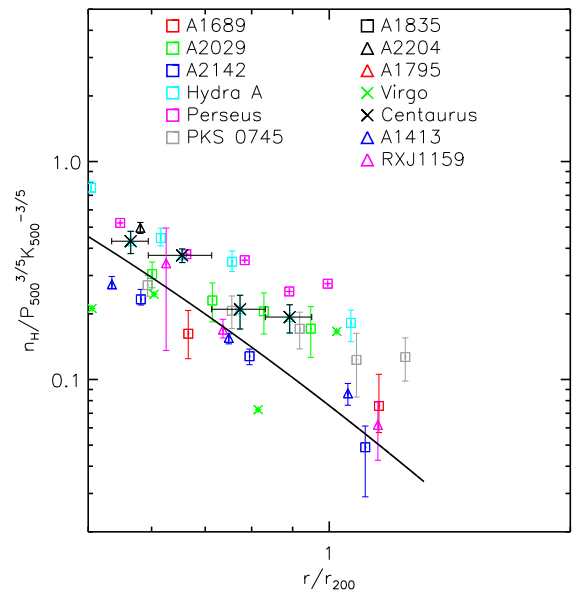, angle=0,
        width=\linewidth}
      }        
      \caption{Comparing the self similar scaled temperatures (top panel) and densities (bottom panel) outside $0.5r_{200}$ with the predictions obtained by assuming that the baseline entropy profile of \citet{Voit2005} and the universal pressure profile of \citet{Arnaud2010} both apply in the outskirts. } 
      \label{compareTandnwithpredictions}
  \end{center}
\end{figure}

Further insight can be obtained by comparing the scaled temperatures and densities measured in the outskirts (outside 0.5$r_{200}$) with the temperature and density profiles that result when both the baseline entropy profile and the universal pressure profile of \citet{Arnaud2010} are assumed to hold in the outskirts. Scaling by the self similar entropy and pressure we can write the expected temperature profile as,
\begin{eqnarray}
\label{ArnaudVoit_T_selfsimilar}
kT^{scaled}(r) &=& (P(r)/P_{500})^{2/5} (K(r)/K_{500})^{3/5} 
\end{eqnarray}
and the expected hydrogen density profile as,
\begin{eqnarray}
\label{ArnaudVoit_n_selfsimilar}
n^{scaled}_{H}(r) &=& (1/1.2) (P(r)/P_{500})^{3/5} (K(r)/K_{500})^{-3/5}
\end{eqnarray}
where $K(r)/K_{500}$ is given by equation \ref{eqn:KR} and $P(r)/P_{500}$ is given by equation \ref{eq:puniv}. Outside $r_{500}$ the $(M_{500} / 3 \times 10^{14}\rm h^{-1}_{70} M_{\odot})^{ \alpha_{\rm P}+\alpha^{\prime}_{\rm P}(x)}$ term in the pressure profile is negligible.  

We scale the observed temperatures by $P_{500}^{-2/5} K_{500}^{-3/5}$ and the observed densities by $P_{500}^{-3/5} K_{500}^{3/5}$, and compare them to equations \ref{ArnaudVoit_T_selfsimilar} and \ref{ArnaudVoit_n_selfsimilar} in Fig. \ref{compareTandnwithpredictions}. No density profiles are presented in \citet{Bonamente2012} (for Abell 1835) or \citet{Bautz2009} (for Abell 1795), so these cannot be shown. 

We see that the temperatures generally agree with the predicted temperature profile out to $r_{200}$, but the temperatures outside this are lower. These low temperatures are the dominant cause of the entropy decrement compared to the baseline profile for the 4 clusters in question (A1689, A2142, A1835 and PKS 0745-191). The good agreement with the predicted temperatures within $r_{200}$ seems to indicate that temperature biases due to cold gas clumps are not significant within $r_{200}$, which can be explained if the gas clumps are not in thermal pressure equilibrium with the surrounding ICM but are mostly confined by ram pressure as they move through the ICM. Outside $r_{200}$ it is possible that the decrease in temperature below the predicted level is at least partly the result of cold gas clumps biasing the temperature low.

Most of the measured densities lie above the predicted density, for which gas clumping may be the cause. This overdensity appears to be the cause of the flattening of the entropy profile for Perseus, Hydra A and Abell 2029, which all have temperatures in agreement with the predicted temperature (i.e. for these clusters the entropy decrement does not occur because the temperatures are too low, but because the densities are too high compared to predictions). By contrast, for A2142 and A1689, the densities are in reasonable agreement with predictions in the outskirts, and the entropy decrement in the outskirts of these clusters appears to be caused by the temperatures being too low compared to predictions. 

\subsection{Comparison with Eckert et al. 2013}

\begin{figure}
  \begin{center}
    \leavevmode

      \epsfig{figure=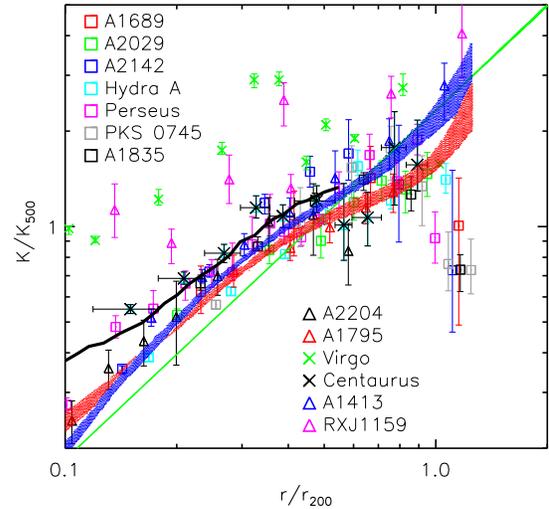, angle=0,
        width=\linewidth}

      \caption{Same as Fig. \ref{entropy_compare} but with the entropy profiles obtained in \citet{Eckert2013a} overplotted. Blue is for CC clusters and red is for NCC clusters (the same colour notation as in \citealt{Eckert2013a}). } 
      \label{E13_compare}
  \end{center}
\end{figure}

Recently, \citet{Eckert2013a} combined Planck pressure profiles and ROSAT PSPC density profiles to derive the entropy profiles for a sample of clusters and claimed that the entropy profiles of cool core clusters agree with the baseline entropy profile outside $r_{200}$. Their results for cool core (CC) and non cool core clusters (NCC) are shown overplotted on the Suzaku results as the blue and red shaded regions respectively in Fig. \ref{E13_compare}. A complete investigation into the discrepancy between the Suzaku results outside $r_{200}$ with those presented in \citet{Eckert2013a} is beyond the scope of this paper, however here we briefly discuss some possible causes.

To obtain the entropy profiles, \citet{Eckert2013a} fitted the Planck pressure and ROSAT density profiles with functional forms, which means that a functional form was assumed for the entropy profile. The priors placed on the entropy profile by the assumption of this functional form, and the degrees of modelling freedom available, were not fully demonstrated. 

As can be seen in Fig C.2 in \citet{Eckert2013a}, the uncertainties on the raw ROSAT density profiles increase dramatically outside $r_{500}$, and the functional forms which are fit to these profiles appear unable to fully explore the errors in the outskirts. This leads to the error envelope of the functional forms (green shaded regions in Fig C.2) significantly underestimating the true density errors outside $r_{500}$. It is unclear how sensitive the fits to the ROSAT density profiles are to the data points outside $r_{500}$, and no statistic is presented to indicate how sensitive the fits are to the outer regions. It is therefore unclear to what extent the functional form fitting is controlled by the central regions where the data quality is higher. 

The underestimate of the errors through the use of functional forms is evident in Fig. 3 in \citet{Eckert2013a} where the temperature profiles derived from the functional form fitting (solid green regions) are compared to those derived using non-parametric deprojection (red triangles). We see that outside $r_{200}$ the use of the functional form causes the temperature errors to be underestimated by at least a factor of 3. This is also evident in the gas mass fraction profiles derived from the same data in Fig. 2 (left panel) of \citet{Eckert2013b}, where the use of functional forms underestimates the errors outside $r_{200}$ compared to the non-parametric method by at least a factor of 3.     

In addition, unlike in \citet{Pratt2010} where the individual entropy profiles for each cluster are displayed to give an indication of the scatter around the mean REXCESS entropy profile, the scatter around the mean entropy profile for each individual cluster is not shown in \citet{Eckert2013a}, so it is unclear what the range of the entropy profiles is. There is for instance a large range in the gas mass fraction profiles for individual clusters shown in Fig. 1 of \citet{Eckert2013b}, which is much larger than the mean profile (green shaded area) they claim from their functional form fitting. The simulations of \citet{Nagai2011} found that the amount of gas clumping can vary significantly between individual clusters, resulting in a large scatter around the median entropy profile in the outskirts. 

\section{Summary}

We have explored the thermodynamic properties of the ICM of the outskirts of the Centaurus cluster, reaching out to 95 percent of the $r_{200}$ value we measure along a strip to the north west. The density profile we have found with Suzaku agrees well with the densities obtained using ROSAT PSPC observations which sample the whole azimuth, indicating that our densities are not biased by looking along only one strip. 

The entropy profile demonstrates the same central excess reported for the REXCESS clusters in \citet{Pratt2010}, and agrees with the baseline entropy profile of \citet{Voit2005} in the outskirts, but is systematically slightly below it (Fig. \ref{T_and_d_profiles}). We find that the pressure profile agrees with the universal pressure profile of \citet{Arnaud2010}, but lies systematically slightly above it in the outskirts. The lower entropies and higher pressures in the outskirts may be the results of gas clumping in the outskirts causing the gas density to be overestimated. This conclusion is supported by the fact that while the temperatures in the outskirts agree with the temperatures obtained by combining the baseline entropy profile and the universal pressure profile, the densities lie above this prediction (Fig. \ref{T_and_d_profiles}).

The gas mass fraction profile we measure rises to the mean cosmic baryon fraction near $r_{200}$, and agrees with the values obtained with Planck in \citet{PlanckV2012} at the largest radius studied. Correcting for the gas clumping needed to make the entropy profile agree perfectly with the baseline profile in the outskirts causes the gas mass fraction to agree with the expected hot gas fraction (which accounts for 12 percent of the baryons being in stars), and also brings the gas mass fraction profile into agreement with the predictions of \citet{Young2011}, (Fig. \ref{gasmassprofile_Young}). Performing this clumping correction also brings the pressure profile into better agreement with the universal pressure profile of \citet{Arnaud2010} (Fig. \ref{pressureprofile}). 

We find that the total mass profile is well described by an NFW profile with $c_{200}$=5.9$^{+1.8}_{-1.4}$, $r_{s}$=191$^{+70}_{-52}$kpc, $r_{200}=1130^{+62.5}_{-54}$kpc$=84^{+4.6}_{-4.0}$arcmins and $M_{200}=1.6^{+0.3}_{-0.2} \times 10^{14}M_{\odot}$. The derived mass is in reasonable agreement with the $M_{200}-T$ relation found in \citet{Arnaud2005} given the scatter around the best fit relation, and lies below the relation, in agreement with the steepening of the $M_{200}-T$ relation away from the self-similar $M \varpropto T^{3/2}$ relation when lower mass clusters are included. 

We have furthered the analysis of the collective properties of the entropy profiles of clusters explored in the outskirts with Suzaku originally reported in \citet{Walker2012_UEP}. When the entropy profiles are scaled by the self-similar entropy, $K_{500}$, we find that the clusters generally demonstrate an excess above the baseline entropy profile of \citet{Voit2005} within $r_{500}$, as has been found for the REXCESS clusters in \citet{Pratt2010} and for a sample of groups in \citet{Sun2009}. The entropy profiles then flatten and agree with the baseline entropy profile at around $r_{500}$. However outside $r_{500}$ the entropy profiles tend to lie below the baseline entropy profile, indicating that non-gravitational processes are present outside $r_{500}$ which lower the entropy. One possibility is that gas clumping is responsible, (as has been shown in the simulations of \citealt{Nagai2011}), and the overdensities required to bring the entropies into agreement with the baseline entropy profile are calculated in Fig. \ref{clumping}. 

We have compared the scaled temperature and density profiles in the outskirts of these clusters (outside 0.5$r_{200}$) with the profiles obtained by assuming both the baseline entropy profile of \citet{Voit2005} and the universal pressure profile of \citet{Arnaud2010} (Fig. \ref{compareTandnwithpredictions}). The temperatures agree with this prediction in the range 0.5-1.0$r_{200}$, however the 4 measurements which have been obtained outside $r_{200}$ (for PKS 0745-191, Abell 1835, Abell 2142 and Abell 1689) all lie below the predicted temperature, causing the entropy to be lower than the baseline level. The density profiles tend to lie above the predicted level outside 0.6$r_{200}$, possibly as a result of gas clumping, which acts to reduce the measured entropy.

\label{summary}

\section*{Acknowledgements}

SAW is supported by STFC, and ACF thanks the Royal Society. This research has used data from the $Suzaku$
telescope, a joint mission between JAXA and NASA.

\bibliographystyle{mn2e}
\bibliography{Centaurus_paper}

\appendix
\section[]{}
\label{sec:appendix}

\begin{figure*}
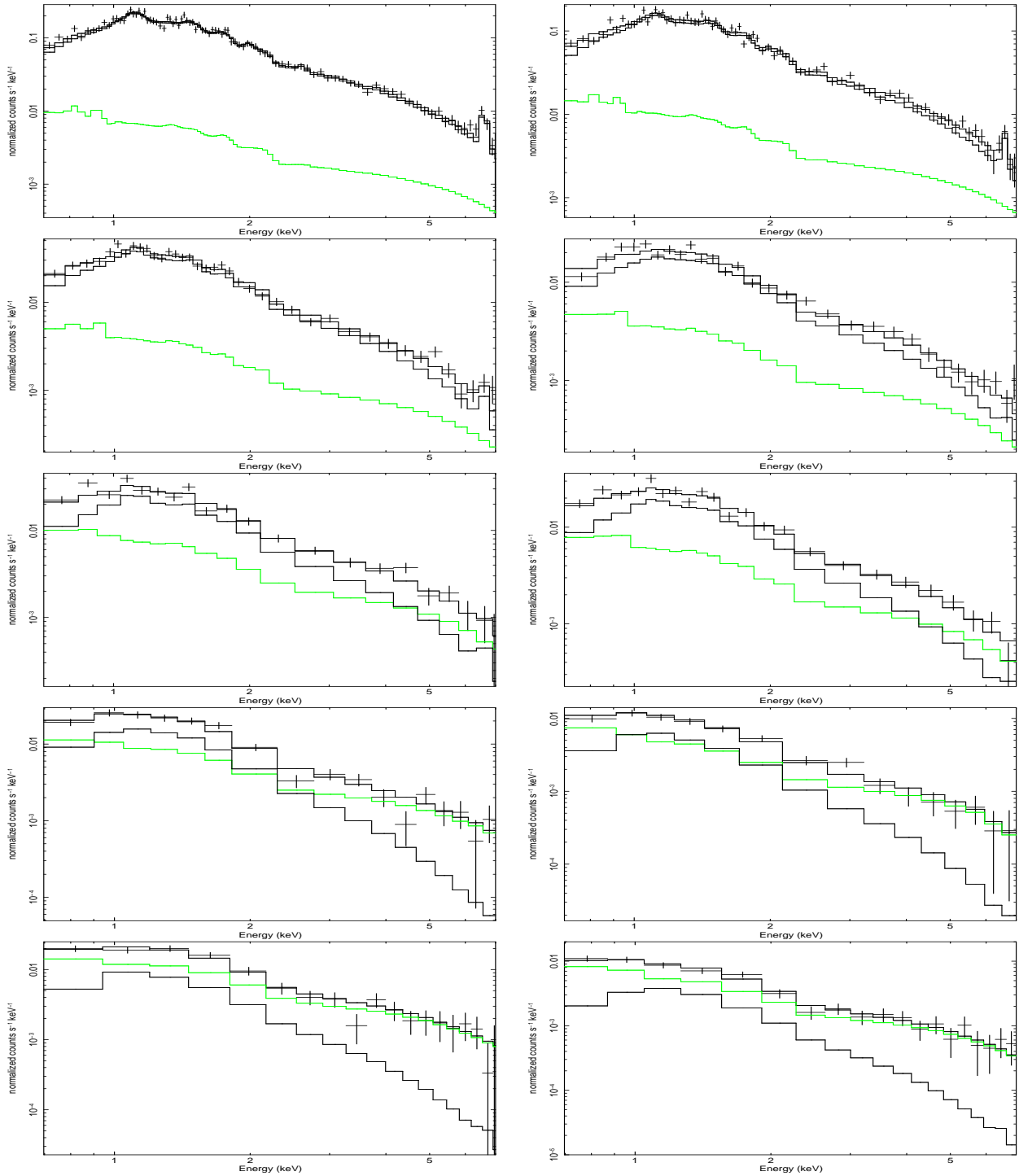

  \begin{center}
     \hbox{\epsfig{figure=FigA1.ps, height=1.0\columnwidth,width=0.45\columnwidth, angle=-90 }
     \epsfig{figure=FigA2.ps, height=1.0\columnwidth,width=0.45\columnwidth, angle=-90 }
     }
      \hbox{\epsfig{figure=FigA3.ps, height=1.0\columnwidth,width=0.45\columnwidth, angle=-90 }
     \epsfig{figure=FigA4.ps, height=1.0\columnwidth,width=0.45\columnwidth, angle=-90 }
     }    
          \hbox{\epsfig{figure=FigA5.ps, height=1.0\columnwidth,width=0.45\columnwidth, angle=-90 }
     \epsfig{figure=FigA6.ps, height=1.0\columnwidth,width=0.45\columnwidth, angle=-90 }
     }   
            \hbox{\epsfig{figure=FigA7.ps, height=1.0\columnwidth,width=0.45\columnwidth, angle=-90 }
     \epsfig{figure=FigA8.ps, height=1.0\columnwidth,width=0.45\columnwidth, angle=-90 }
     } 
     
                 \hbox{\epsfig{figure=FigA9.ps, height=1.0\columnwidth,width=0.45\columnwidth, angle=-90 }
     \epsfig{figure=FigA10.ps, height=1.0\columnwidth,width=0.45\columnwidth, angle=-90 }
     }

      \caption{Spectral fitting for the cluster emission. In reading order we show the spectra from the annuli between 10$'$-15$'$, 15$'$-20$'$, 20$'$-25$'$, 25$'$-30$'$, 30$'$-35$'$, 35$'$-45$'$, 45$'$-50$'$, 50$'$-60$'$, 60$'$-70$'$ and 70$'$-80$'$. The data from each detector and each editing mode were fitted simultaneously and are added here for display purposes only. The black lines through the points represent the best fits (background plus cluster emission), while the lower black line shows the cluster emission. The green line shows the X-ray background level. }
      \label{cluster_spectra}
  \end{center}
\end{figure*}

\end{document}